\documentclass[opre,nonblindrev]{informs3}
\usepackage{amsmath}
\usepackage{amssymb}
\usepackage[numbers]{natbib}

\usepackage{natbib}
 \bibpunct[, ]{(}{)}{,}{a}{}{,}%
 \def\bibfont{\small}%

 \TheoremsNumberedThrough
 \ECRepeatTheorems
 \EquationsNumberedThrough

\newcommand{\paranth}[1]{\left(#1\right)}
\newcommand{\bracket}[1]{\left[#1\right]}
\newcommand{\curly}[1]{\left\{#1\right\}}

\graphicspath{{./figures/}}

\TITLE{Penetrating a Social Network:\\ The Follow-back Problem}
\RUNTITLE{The Follow-Back Problem}

\date{\today}

\RUNAUTHOR{Que, Rajagopalan, and Zaman}
\ARTICLEAUTHORS{%
\AUTHOR{Fanyu Que}
\AFF{%
Boston College \\ 
Boston, MA \\ 
\EMAIL{quef@bc.edu }
}
\AUTHOR{Krishnan Rajagopalan}
\AFF{%
Operations Research Center, Massachusetts Institute of Technology\\
Cambridge, MA  02139 \\
\EMAIL{krishraj@mit.edu}
}
\AUTHOR{Tauhid Zaman}
\AFF{%
Sloan School of Management, Massachusetts
Institute of Technology \\ 
77 Massachusetts Ave. \\
Cambridge, MA 02139 \\ 
\EMAIL{zlisto@mit.edu}
}

}

\ABSTRACT{
	Modern threats have emerged from the prevalence of social networks.  Hostile actors, such as extremist groups or foreign governments, utilize these networks to run propaganda campaigns with different aims.  For extremists, these campaigns
	are designed for recruiting new members or inciting violence.  For foreign governments, the aim
	may be to create instability in rival nations.  Proper social network
	counter-measures are needed to combat these threats.  Here we present one important counter-measure: penetrating social networks.  This means making target users connect with or \emph{follow} agents deployed in the social network.  Once such connections are established with the targets, the agents can influence them by sharing content which counters the influence campaign.	In this work we study how to penetrate a social network, which we call
	the \emph{follow-back problem}.  The goal here is to find a policy that maximizes the number of targets that follow the agent.
	
We conduct an empirical study to understand what behavioral and network features affect the probability of a target following an agent.  We find that the degree of the target and the size of the mutual neighborhood of the agent and target in the network affect this probability. Based on our empirical findings, we then propose a model for targets following an agent. Using this model, we solve the follow-back problem exactly on directed acyclic graphs and derive a closed form expression for the expected number of follows an agent receives under the optimal policy.  We then formulate the follow-back problem on an arbitrary graph as an integer program.  To evaluate our integer program based policies, we conduct simulations on real social network topologies in Twitter.  We find  that our polices result in more effective
network penetration, with significant increases in the expected number of targets that follow the agent. 
   }


\begin{document}

\maketitle

\section{Introduction}

The  prevalence of online social networks has raised many new security
threats.  By penetrating social networks, malicious actors have the potential
to  run massive influence campaigns on a population and disrupt societies.
For instance, a Turkish hacking group gained control of several
Twitter accounts that are followed by U.S. President Donald Trump \citep{ref:turkish_hack}.  The hackers sent private direct messages from these accounts to President Trump containing a link, which had it been clicked, would have provided the hackers the password for his Twitter account.  Control of the U.S. President's Twitter account would allow the hackers to potentially trigger conflicts and create chaos.  It would have been a grave national security threat.  

While this hacking case is an extreme example, it does suggest the dangers of social network penetration.  Hacking is one way to penetrate, but there are other less
invasive, yet still effective methods.  For instance,  instead of directly hacking President Trump's followers, the hackers could simply interact with these followers in order to be followed by them and gain influence over them.
Then, if President Trump's followers shared any content these accounts posted, he would see it, and potentially be influenced by it.  Or  he could follow these accounts directly, giving them a direct line to the U.S. President.  Because Donald Trump is an avid social media user, these scenarios are very plausible.  In fact, he is known to share content posted by his followers occasionally on Twitter \citep{ref:cnngif,ref:cnntraingif,ref:trumpchicago}.  In this manner, a team of hackers could slowly over longer periods of time shift the opinion of the U.S. President, or other influential individuals, in order to align with their agenda.

Aside from the U.S. President, U.S. citizens could also be targets of such 
soft penetration attacks.  There have been multiple reports alleging that foreign actors attempted to penetrate U.S. social networks in order to manipulate elections \citep{ref:russianbots}.  There have been studies showing similar operations in European elections \citep{ferrara2017disinformation}.  The perpetrators created fake or ``bot'' accounts
which shared  politically polarizing content, much of it fake news, in order to amplify it and extend its reach, or directly interacted  with humans to promote their agenda \citep{ref:russianbots1}.    While no one knows exactly how many people were impacted by these influence campaigns, it has still become a concern for the U.S. government.  Members of Congress have not been satisfied with the response of major  social networks \citep{ref:russianbots_govtresponse} and have asked them to take
actions to prevent future interference in the U.S. democratic process by foreign actors   \citep{ref:russianbots_feinstein}.

Penetrating social networks is also a tool that can also be used to counter
violent extremists in social networks.  This topic has gained the interest of national defense policy makers and strategists. The U.S. State Department  launched a handle on Twitter, @ThinkAgain\_DOS, as a means to insert counter-propaganda content into networks of jihadist sympathizers. The profile engaged with many prominent members of the Islamic State in Iraq and Syria (ISIS), a known extremist group, on Twitter in the fall of 2014, but was widely criticized as being unsuccessful \citep{think_again}. American leadership is still trying to understand how to combat extremist recruitment over social media. In January of 2016, The White House held a summit with executives from some of the leading American technology firms, including Facebook and Twitter, to help generate strategies to counter extremist recruitment online \citep{ref:WhiteHouseSiliconValley}. 

Having effective social media based counter-measures
is essential to stopping foreign based influence campaigns or 
violent extremists.
Military leaders have been tasked with similar challenges before, but social media is a relatively new operating environment. The United States Department of Defense uses Military Information Support (MIS) units to penetrate contested information environments. Military Information Support Operations (MISO) are the conveying of specific information to target foreign audiences in order to sway their perceptions and ultimately influence their actions in support of national objectives \citep{ref:MISO}. When an adversary is attempting to use propaganda to influence a target audience over social media, MIS units may be tasked with inserting counter-propaganda information in the network.  As social media platforms have changed the way information is spread, MIS units require tools that will allow them to operate in this new, increasingly important space.  

One problem confronting MIS units is ensuring that their counter-propaganda efforts can penetrate  social networks.  On Twitter, for instance, a user will generally only see the content posted by users who he is connected with or \emph{follows}. MIS units must form these connections in order to operate in online social networks. Therefore, a key capability needed is a systematic way to penetrate these networks by forming the appropriate connections.

\subsection{Our Contributions}
In this work we present an operational capability for penetrating a social network.
We refer to our modeling framework as the \emph{follow-back problem}, where the goal is to find the most effective way for an agent to interact with users in a social network in order to maximize the number of social connections.  The name is based on terminology from the social network Twitter, where when a user forms such a connection he is said to become a \emph{follower}. If an agent is able to have a set of target users
follow him, then it has effectively penetrated their social network and can then begin
influencing these target users with its content. 

The follow-back problem uses a model for user behavior in social networks, in particular, what features affect the probability of being followed.  To develop this model, we begin by  conducting a field experiment in Twitter where online agents interact with users in order to gain followers.  Our experimental findings show that the follow probability is affected by the number of friends (users following the targeted user) and followers of the targeted user.  It is also affected by the number of friends of the targeted user that follow the agent, a feature we refer to as the \emph{overlap}.  In particular, we find that the follow probability is larger if the targeted user has many friends, few followers, and a large overlap with the agent.  

We propose a simple model for the follow probability based on our empirical results and use this model as the basis for our analysis of the follow-back problem on different graph structures.  We are able to calculate the optimal interaction policy for directed acyclic graphs (DAGs) and provide an expression for the expected number of targets that follow the agent under this policy. For graphs with arbitrary cycles, we use the results for DAGs as the basis for an integer programming formulation which finds good policies.

We test the resulting  policies using simulations on
multiple real Twitter network topologies, including the network of U.S. President Donald Trump.
We find that our policies can increase the expected number of follows
over simple policies by an order of magnitude.  We also find that our policies consistently outperform network centrality based policies.  In short, we find that our integer programming based interaction policies allow agents to penetrate social networks much more efficiently.

The remainder of this work is outlined as follows.  We review related literature in Section \ref{sec:LitReview}.  We formally define the follow-back problem in Section \ref{sec:def}.  We then present our empirical analysis of follow probabilities in   Section \ref{sec:experiments}.  We analyze optimal policies for the follow-back problem on DAGs in Section \ref{sec:theory}. Then, we formulate the follow-back problem as an integer program for arbitrary graphs in Section \ref{sec:ip}.  We compare the performance of our policies versus other policies in Section \ref{sec:results}.  Finally, we discuss potential implications of our work and some avenues of future research in Section \ref{sec:conclusion}.

\section{Literature Review} \label{sec:LitReview}

There are several lines of research related to our work.  
First, there is work on empirical studies of influence in social networks. 
Second, there is the body of work on the sociological notion of triadic closure and its manifestation in online
social networks. 
Third, there is a sizable body of work on theoretical analysis of influence maximization in social networks.  
Finally, there is the broad area of network centrality measures.

\subsection{Empirical Measurement of Influence in Social Networks}

Many researchers have studied what factors affect influence in social networks.  Influence can be roughly defined as the ability to cause an individual to adopt a behavior.  The ability of social influence to affect health behavior was demonstrated in \cite{christakis2007spread} and \cite{christakis2008collective}.  The effect of social network structure  on health behavior was demonstrated in \citep{centola2010spread}.  The use of randomized online experiments to discern causal social influence  was done in \cite{ref:sinanIEEE}.  In \cite{ref:sinanScience} the authors conducted a massive randomized online experiment in Facebook to infer what types of  users are  influential and  susceptible.  It was found that influence and susceptibility depended on demographic characteristics as well as network structure.   \cite{ref:miacha} analyze the influence of users in Twitter based upon number of retweets and number of followers.  \cite{kleinbergFacebook} studied structural features of networks that affect influence in Facebook.  

\subsection{Triadic Closure}
A key element to the follow-back problem is the sociological notion
known as \emph{triadic closure} \citep{granovetter1977strength}.  This notion
is closely related to the overlap measure we study in Section \ref{sec:overlap}.
  The effect of triadic closure on influence was demonstrated empirically  in \cite{kleinbergFacebook} and \cite{ref:nagle}. 
  In \cite{ref:socialbotnet} the authors found that triadic closure had an effect
  on an agent's ability to form connections with new users in Facebook.  While triadic closure was initially defined for undirected graphs, there have been
  extensions developed for directed graphs \citep{ref:kleinberg_directed} which we 
  utilize in this work. 

\subsection{Influence Maximization in Social Networks}
Related to the follow-back problem is the influence maximization problem where  the goal is  to maximize the spread of a piece of information through a social network by seeding a select set of users. 
One of the first theoretical studies of the influence maximization problem was in \cite{kempe2003maximizing} where  it was shown that the problem was sub-modular, so a greedy algorithm would have good performance.  This work led to subsequent variations of the problem and different algorithmic solutions \citep{kempe2005influential, chen2010scalable, chen2009efficient}. Each of these works exploited the sub-modular structure of the problem to obtain efficient algorithms.

\subsection{Network Centrality}
Many problems in graphs can be solved through the development of functions known as \emph{network centralities} which map vertices to real numbers.  These functions quantify how central a vertex is to the problem at hand, with the definition of centrality being set by the application.  In our work we compare our integer programming based policies to policies based on network centralities. One of the simplest is distance centrality which measures how close a vertex is to all other vertices \citep{ref:dc}.  A related measure is betweenness centrality, which measures how many paths pass through a vertex \citep{ref:bc}.   Rumor centrality was developed to find the source of a rumor that spreads in a graph \citep{shah2011rumors}.  Eigenvector centrality measures how important a vertex is by how important its neighbors are \citep{ref:ec1}.  This centrality measure is closely related to the PageRank \citep{ref:pagerank} and HITS \citep{ref:hits} centrality measures.  


\section{Follow-back Problem Definition} \label{sec:def}
We consider a directed social network where an edge $(u, v)$ pointing from $u$ to $v$  means that $u$ is \emph{followed} by $v$. In a social network such as Twitter this means that any content $u$ posts will be visible to $v$. Therefore, the directed edges indicate the flow of information in the network.  We begin with a set of target users $T$ in a social network that we want to follow an agent $a$.  To gain trust with the targets, the agent will interact with users that the targets follow, also known as the targets' \emph{friends}, though the agent can in theory interact with any users in the network.  We denote the agent, targets, and any other users the agents interact with as the set $V$.  From this we have that $T\subset V$. In the social network these individuals form what we refer to as the \emph{friends graph}, denoted by  $G_0 = (V,E_0)$ where $E_0$ is the initial edge set. This edge set will change as the agent interacts with the users in $V$.
  
\begin{figure} 
	\centering
	\includegraphics[scale = .9]{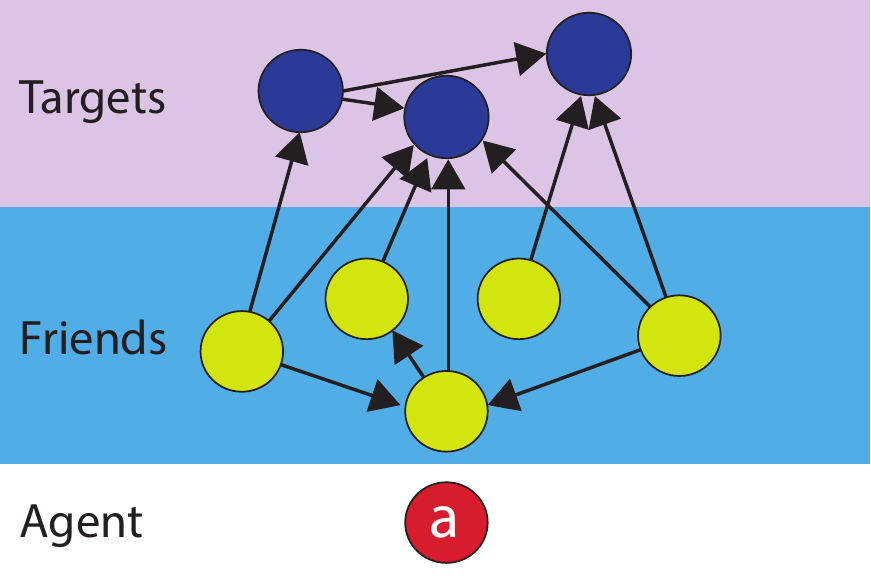}
	\caption{Illustration of the initial friends graph in the follow-back problem.  The agent $a$ wants to maximize its expected number of follows from the targets by interacting with all vertices in the friends graph.} 
	\label{fig:followback_illustration} 
\end{figure}

Each vertex $v\in V$ has a set of features $F_v$ associated with it and we define the set of all vertex features as $\mathbf F = \bigcup_{v\in V}F_v$. 
We will assume that the agent initially has no edge with any other vertex in $V$.  
The goal in the follow-back problem is for the agent to maximize the expected number of follows it gets from the target vertices in $T$ by interacting with vertices in $V$. We illustrate the setting of the follow-back problem in Figure \ref{fig:followback_illustration}.

At each time $n$, the agent $a$ can interact with a single vertex
in $V$.  The vertex chosen by the agent at time $n$ is denoted $u_n$.  The interaction of the agent with a target vertex is determined by the social network.  For instance, in Twitter, these interactions include following, retweeting, and mentioning.  We require that $u_n \neq u_i$ for any $i < n$, meaning that the agent can only interact with each vertex once.  This is true for actions such as following, which can only be done once. We define the interaction policy of the agent as the vertex sequence $\pi = \curly{u_1,u_2,...}$ along with the sequence of interactions $ x =\curly{x_{u_1},x_{u_2},...}$.  Here $x_{u_i}$ specifies the type of  interaction  with $u_i$. When the agent interacts with a vertex $v$, either $v$ will follow it or not.  We let the random variable $X_{v}$ be one if  $v$ ends up following $a$, otherwise $X_{v}$ is zero. We refer to the probability of the agent gaining a follow from a vertex as the \emph{follow probability}.  Gaining followers does not change
the vertex set, but it does change the
edge set $E_n$. If the agent's action  gains a follow from vertex $u_n$, then a new edge is created from $u_n$ to the agent $a$ and we update the graph to $G_n = (V,E_n)$ where $E_n = E_{n-1}\bigcup (a,u_n)$. 

In the follow-back problem, the agent selects the policy $(\pi,x)$ to maximize the expected number of target users  that follow-back.  That is,  the agent chooses a policy $(\pi,x)$ to solve the following problem:
\begin{equation}
\max_{(\pi,x)} \sum_{v \in T}\mathbf E\bracket{X_v|\pi,x,G_0,\mathbf F}.
\end{equation}
Key to solving this problem is understanding how the follow probabilities
are affected by the interaction type, features, and graph structure.   We accomplish this with empirical analysis, which we discuss next in Section \ref{sec:experiments}.


\section{Empirical Analysis of Interactions in Social Networks}\label{sec:experiments}
 Understanding what affects the follow probability of users in a social network  is a key input to the follow-back problem.
To determine the structure of  follow probabilities we require data on interactions in social networks.
We obtain such data by conducting experiments involving  Twitter accounts to which we were given access.
The accounts belonged to six Moroccan artists who wished to connect with Twitter users in order to promote their artwork.  The artist accounts would interact with Twitter users using different types of interactions, and we recorded the results of the interactions after one week. We were able to obtain data for over 6,000 Twitter interactions.  We now discuss our empirical findings in detail.

\subsection{Empirical Setup}
We conducted a series of experiments to understand different aspects of interactions in social networks.
As mentioned above, we utilized six different Twitter accounts from Moroccan artists which acted as our agents.  We refer to anyone the agents interacted with as a target (this is different from the definition of targets for the follow-back problem).
  We conducted three different experiments. The first experiment was designed to understand the effect of the different types of interactions on the follow probability.  The second experiment aimed to understand how features such as  friend count and follower count affect the follow probabilities.  The third experiment looked at more complex graph features, such as the number of the agent's followers who are also friends of the target, which we refer to as the \emph{overlap}.

We began by creating Twitter accounts for the agents.   The six different  agents we used have different names and profile images, but the content they posted on their Twitter timelines is similar.  This content consisted of images of their artwork, which was similar across the different accounts.  Each agent would interact with a set of targets.  We made sure that each agent/target pair only interacted at most once to avoid the risk of a target's reaction being confounded by multiple interactions.        

For the first two experiments we had the agents use the Twitter search API \citep{ref:twitterSearchAPI} to find targets who have posted tweets with the keywords ``Morocco'' or ``art''.     
The agents then interacted with all of these targets using some form of Twitter interaction: either a retweet, reply, or a follow.  We had the agents perform this search and interact procedure in randomly spaced intervals that were on average two hours apart.  We chose this rate of interaction to make the agents appear more human. Rapid  activity is often the sign of Twitter bots, or automated accounts, which could affect the response of the targets or result in the accounts being suspended by Twitter. The experiments were each run for one week.  We waited one week from when the interaction occurred and checked if there was any sort of response from the targets, such as a follow, a reply, or a retweet.  For the third experiment the procedure was the same, except that to find target users, in addition to using the search API, we also searched for followers of users who had followed the agents during the first two experiments.  Our hypothesis was that these followers would have greater trust in the agents because their friends also followed the agent.

\subsection{Interaction Type} \label{sec:Interaction}
There are three main ways an agent can interact with a target in Twitter.  The first is following.  When the agent follows a target any content the target posts is shown to the agent in its Twitter timeline.    The second interaction is retweeting, which is reposting one of the target's tweets.  The agent's timeline displays the original tweet.  The third interaction is replying, which involves the agent posting a tweet to its timeline that mentions the target.    For all of these interactions, the target is notified by Twitter when it occurs.  

We had one agent account post original content and not interact with other users.  This agent acted  as a control.  The remaining five agents interacted with 150 to 220 different targets over a one week period.  Each agent was assigned a specific type of interaction.  These interactions were following, retweeting, replying, retweeting and following, and replying and following.  Agents who replied used a fixed set of messages such as ``I like that'' or ``Nice''.

\begin{figure} 
	\centering
	\includegraphics[scale = 0.55]{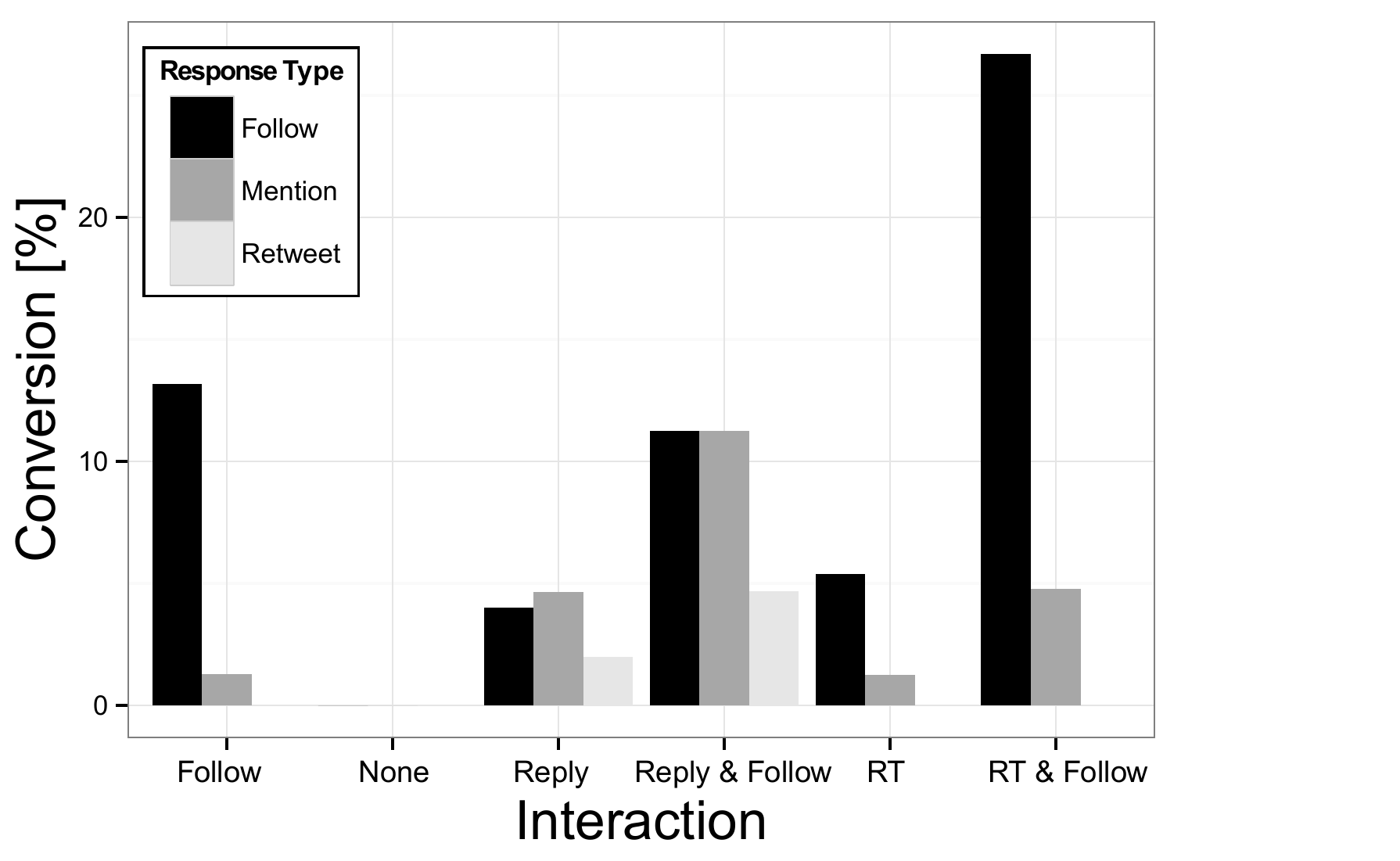}
	\caption{Conversion rate for different agent interaction and target response combinations (RT stands for retweet).}  
	\label{fig:InteractionTypes}
\end{figure}

The target can respond to the agent's interaction by mentioning, retweeting, or following the agent.  If the agent's interaction produces a response from the target, we refer to this as a \emph{conversion}.  We show the conversion rate for each combination of agent interaction and target response in Figure \ref{fig:InteractionTypes}.  The control agent that did not interact with anyone is referred to as ``None'' in the figure.  This agent did not receive any form of target response. 

We focus on the agent being followed by the target, since this is the response of interest for the follow-back problem.  We see in Figure \ref{fig:InteractionTypes} that the most effective interaction to obtain follows is following and retweeting the target.  The conversion rate is nearly 30\%.   Retweeting alone gives a conversion rate of of approximately 5\%. and following alone gives an approximate conversion rate of 14\%.  
The combined effect of these two interactions is greater than their separate individual effects.   Based on this observation, for our remaining experiments we have all agents follow and retweet each target.

\subsection{Friend and Follower Counts}\label{sec:follow_count}
Our next experiment aimed to understand how the friend and follower counts of both the target and agent impacted the follow probability.  As mentioned earlier,  the agents only follow and retweet the targets.  Before starting the experiment we used an online service to obtain followers for the agents.  This allowed for sufficient variance in their follower counts, with one agent receiving as many as 10,000 followers.  These follower accounts appeared to be automated bot accounts.  However, our interest was in changing the follower count displayed on the agent's profile, not in the quality of the followers themselves.  We wanted to see if having a higher follower count would make the agent seem more important and increase the follow probability.  We were able to measure over 2,000 interactions for the agents.

\begin{figure} [t]
	\centering
	\includegraphics[scale = 0.45]{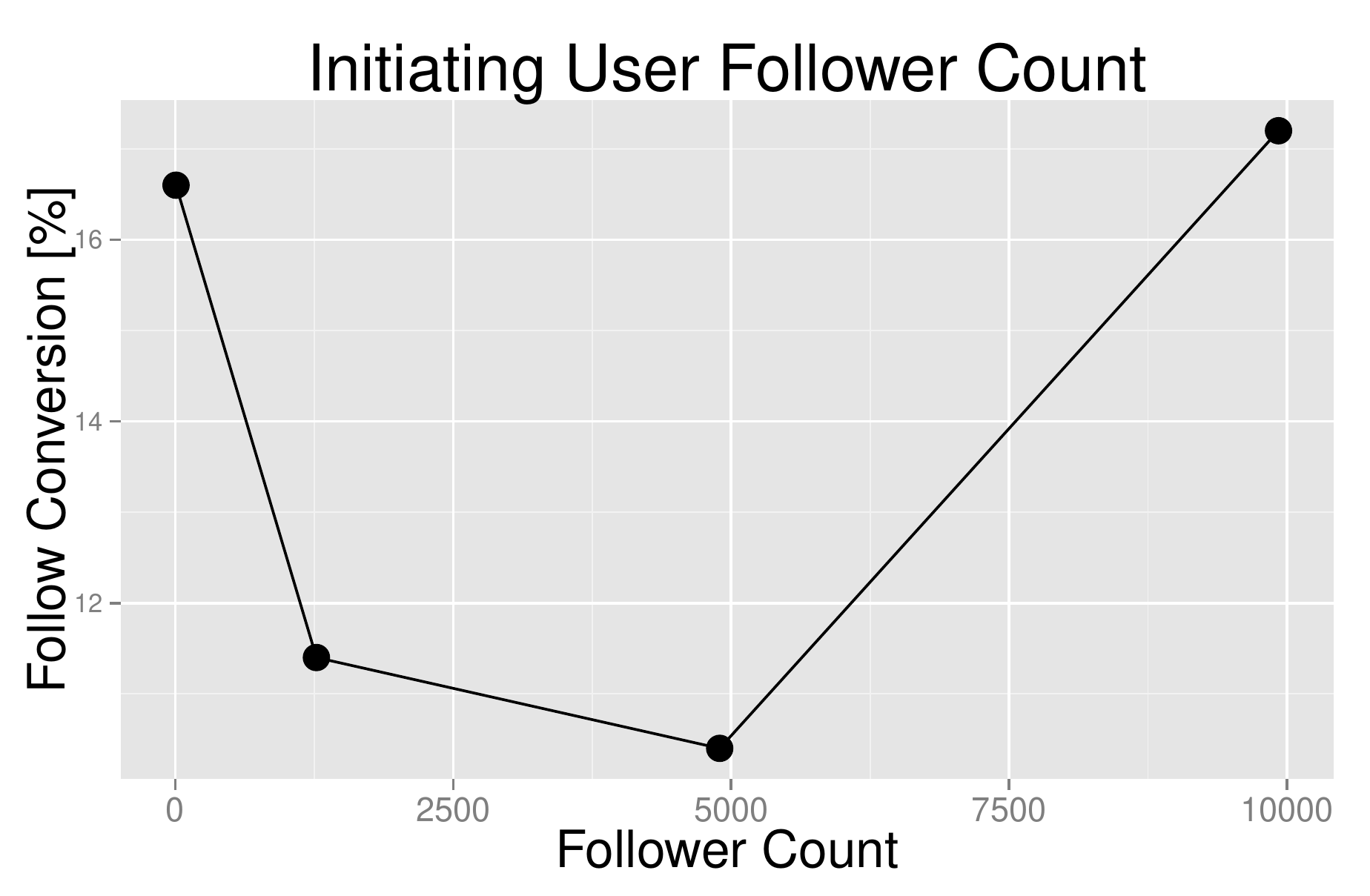}
	\caption{Plot of the follow conversion rate versus agent follower count.  The horizontal axis shows the follower counts of four agents. The vertical axis shows the percentage of targets that followed each agent.}  
	\label{fig:agentfollower}
\end{figure}

We first examine the impact of the agent follower count on the follow probability.  In Figure \ref{fig:agentfollower} we plot the conversion rate for follows versus the follower count of the agent.  The results indicate that the follower count of the agent does not have a significant impact on the conversion rate for follower counts up to 10,000. This is somewhat counter-intuitive as we would expect users to be more likely to follow someone with a higher follower count, as this is a symbol of status on Twitter.   Our data suggest that it is not clear if the agent's follower count has any effect on the follow probability up to 10,000 followers.  There may be an effect for higher follower counts, but we were not able to measure this. 

\begin{figure} [t]
	\centering
	\includegraphics[scale = 0.4]{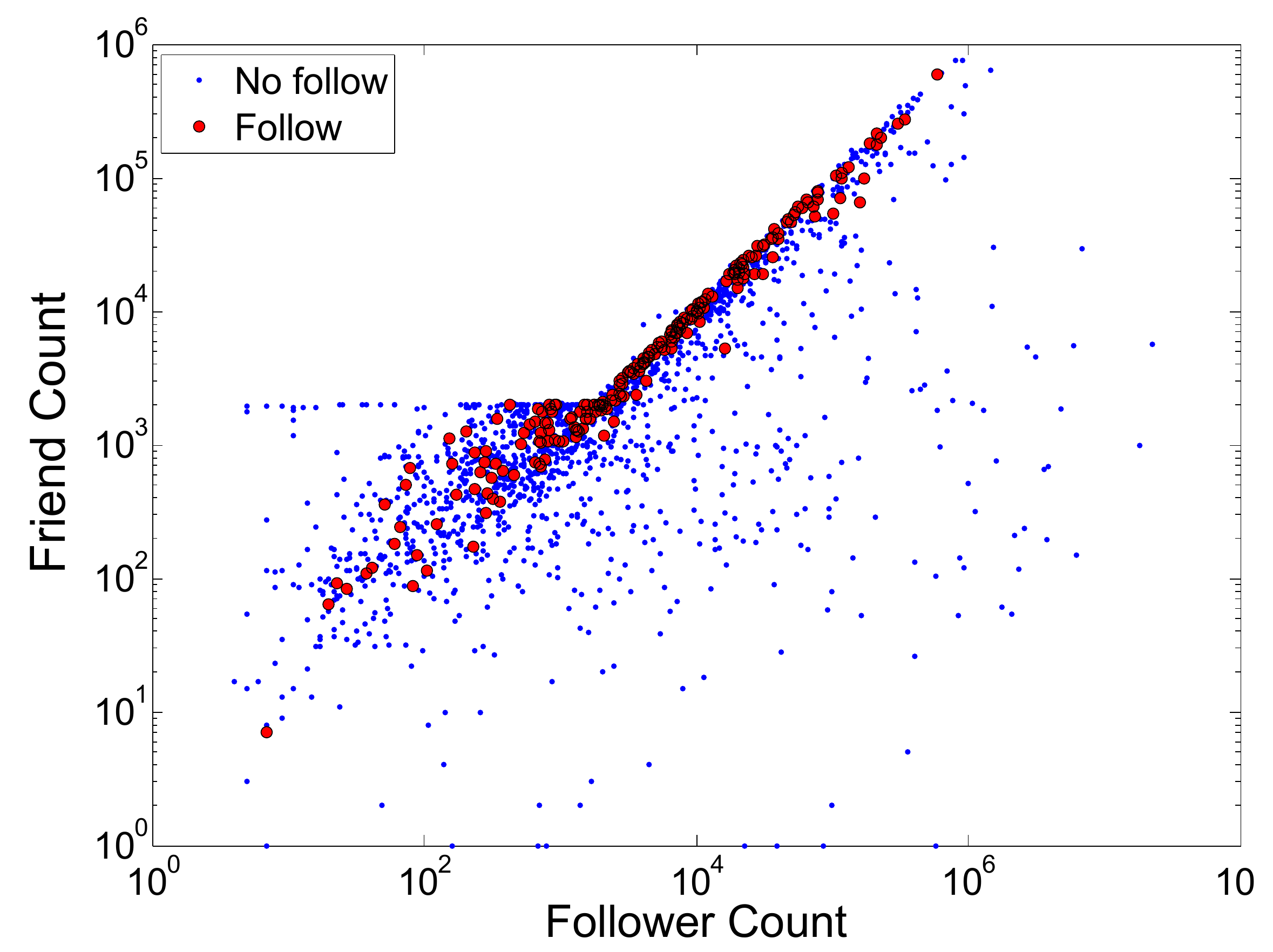}
	\caption{Plot of the follower count versus the friend count for each target.  The markers indicate which targets followed the agent and which did not.}  
	\label{fig:friendfollower}
\end{figure}

The follower and friend counts of the target had a clearer relationship to the follow probability. We plot in Figure \ref{fig:friendfollower} the friend and follower count of each target.  We use different markers for those that followed the agent and those that did not.  A clear pattern is seen here.  Targets with lower follower counts and higher friend counts tended to follow the agents more frequently.  This makes sense intuitively as we would assume that user's with higher friend to follower ratios have a propensity to follow others on Twitter.  

We next perform a formal statistical analysis of the follow probability using a logistic regression model.  We denote the follow probability of an agent/target interaction  as $p(F)$ where $F$ is a vector of features associated with the interaction.   Under our logistic regression model the follow probability for an interaction is 
\begin{equation}
p(F) = \frac{e^{\sum_{k=1}^{d} \beta_kF_{k}}}
{1+e^{\sum_{k=1}^{d} \beta_kF_{k}}}\label{eq:logistic},
\end{equation}
where we have assumed that the set $F$ has $d$ elements. Furthermore, we assume each interaction is an independent event. 

Our regression model used the logarithm (base 10) of the friend and follower counts of the agent and targets as features.  The resulting regression coefficients are shown in Table \ref{table: followerfriendreg} and support what was seen in the figures.  The follower and friend counts for the target were significant while the friend and follower counts of the agents were not.  Also, targets with many friends and few followers were more likely to follow the agent.

\begin{table} 
	\centering
	\caption{Regression coefficients for the follow probability. 
		The features are the log transformed friend and follower count of the target and agent.} \label{table: followerfriendreg}
	\begin{tabular}{|l||c||c|}
		\hline
		Feature & Coefficient& p-value\\ \hline 
		Intercept & -2.641*** & $<2\times 10^{-16}$ \\ \hline
		$\log_{10}$(Target Friend Count) & 3.41*** & $6.28\times 10^{-10}$ \\ \hline
		$\log_{10}$(Target Follower Count) & -2.0*** & $1.03\times 10^{-5}$ \\ \hline
		$\log_{10}$(Agent Follower Count) & 1.5 & 0.29\\ \hline
		$\log_{10}$(Agent Friend Count) & 0.30 & 0.76 \\ \hline
	\end{tabular}
	\\
	The significance codes are 
	$^{***}: 0.001$, $^{**}: 0.01$, and $^*: 0.05$.
\end{table}

\subsection{Overlap}\label{sec:overlap}
\begin{figure} 
	\centering
	\includegraphics[scale = 0.6]{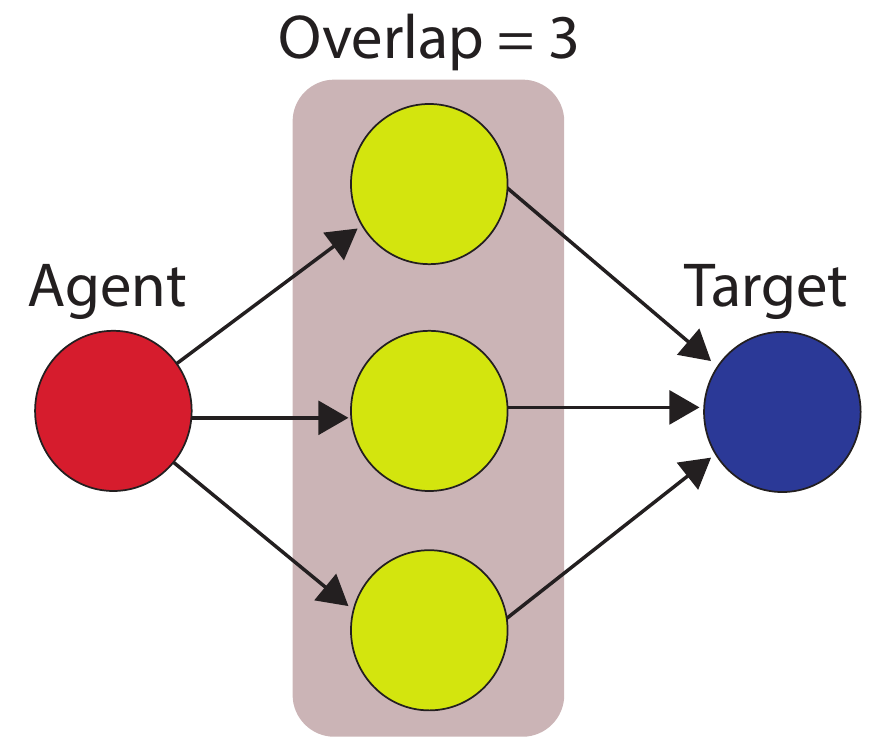}
	\caption{Illustration of the overlap feature between the agent and target vertex.  Overlap is defined as the number of users the target follows who also follow the agent. The overlap in the figure is three.}  
	\label{fig:overlap}
\end{figure}

Our final experiment looked at the overlap feature, which is related to the concept of triadic closure from social network theory.  Triadic closure is the property that if two individuals $A$ and $B$ both have strong ties to $C$, then it is likely that some form of tie exists between $A$ and $B$.  Strong triadic closure states that, given this mutual connection with $C$, a  weak tie between $A$ and $B$ \emph{must} exist \citep{ref:kleinbergBook}.    Triadic closure is a term usually applied to undirected graphs. For directed graphs, \cite{ref:kleinberg_directed} propose related notion known as \emph{directed closure}. We use this concept to define a  feature we call the \emph{overlap} of two vertices.  Formally, the overlap of a vertex $v$ with another vertex $u$ in a graph is equal to the number of directed two-hop paths from $v$ to $u$ in the graph.  In Twitter terms, user $v$ has an overlap with user $u$ equal to the number of friends of $u$ that follow $v$.  We illustrate this overlap feature in Figure \ref{fig:overlap}.  The intuition behind this definition is based on the flow of information.  In the case of Twitter, content posted by a user can be seen by his followers.   Therefore, the target could see some of the agent's content if it is retweeted by friends of the target that follow the agent. Seeing this content may make the target more interested in the agent and result in an increased follow probability.

\begin{figure} 
	\centering
	\includegraphics[scale = 0.4]{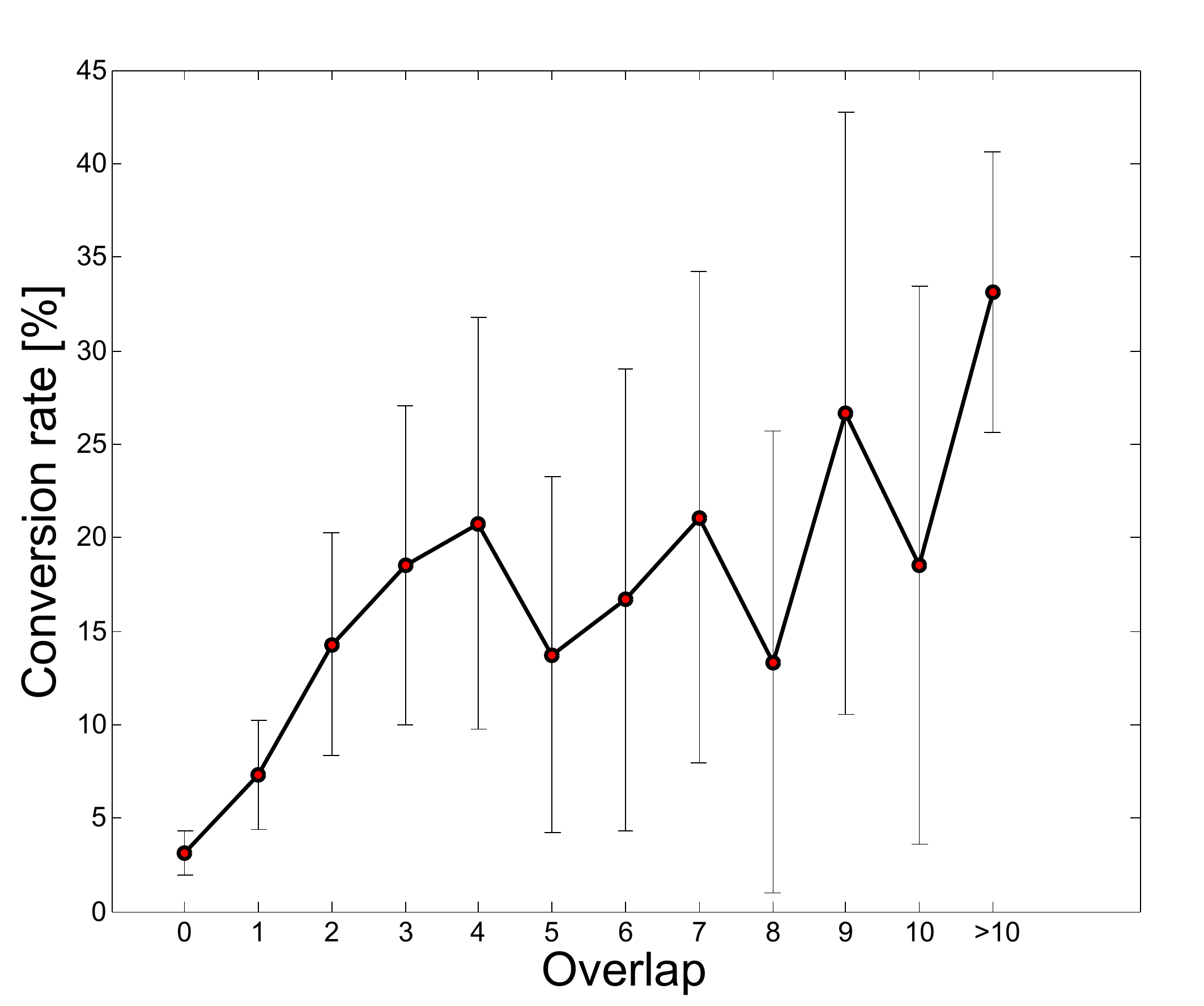}
	\caption{Plot of the follow conversion rate versus overlap. The red circles are the median and the error bars are 95\% confidence intervals.}  
	\label{fig:overlap_plot}
\end{figure}

We had the agents select targets by using the Twitter search API and also by choosing the followers of users who had previously followed the agent.  This way we were able to obtain targets with non-zero overlap.  The agents retweeted and followed each target.  We plot the follow conversion rate versus the overlap in Figure \ref{fig:overlap_plot}.  As can be seen, the conversion rate increases as the overlap increases.  The increase is not monotonic, but this is likely due to not having sufficient data points for different overlap values, as seen by the 95\% confidence intervals in the figure. 
 
\begin{figure} [t]
	\centering
	\includegraphics[scale = 0.38]{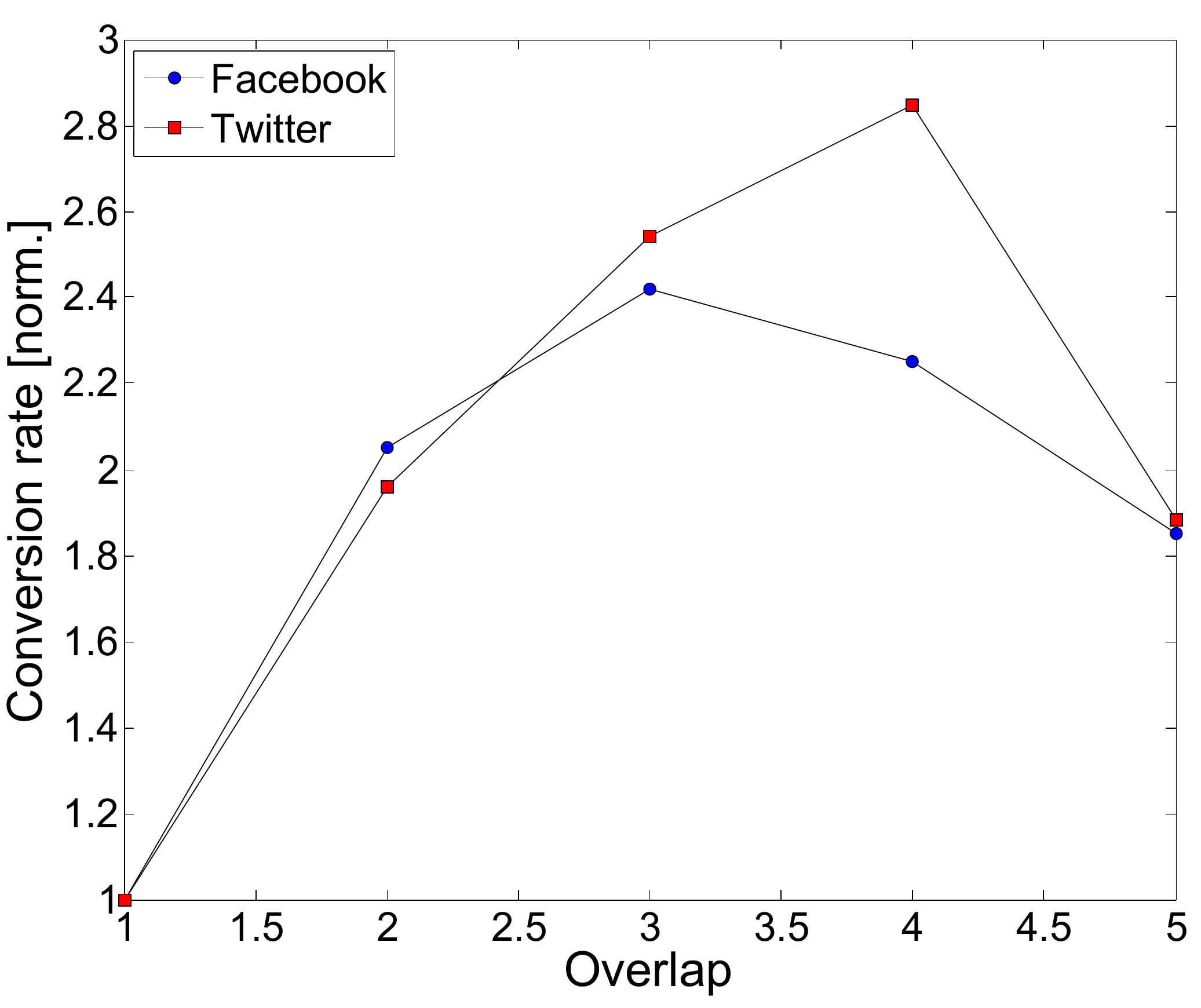}
	\caption{Plot of the normalized conversion rate versus overlap for follows in our Twitter experiment
		 and Facebook signups from  \cite{kleinbergFacebook}. } 
	\label{fig:overlap_plot_facebook} 
\end{figure}

An interesting result is found if we compare our measured conversion rates with those from \cite{kleinbergFacebook}.  In that work, the conversion rate was for users who were invited to Facebook that actually joined, and the equivalent of our overlap feature was the number of email invitations the user received from distinct people. 
Here one can view the agent as Facebook, the user as the target, and the people that invite the user as those that follow Facebook and are followed by the user.  We plot the conversion rate versus overlap from our Twitter experiment and the Facebook experiment of \cite{kleinbergFacebook} in Figure \ref{fig:overlap_plot_facebook}.  To compare results, we normalize the conversion rates so that an overlap of one has conversion rate equal to one (this is the data provided in \cite{kleinbergFacebook}).  Remarkably, we see that the two curves are almost equal, with slight differences for an overlap of four.  The normalized conversion rate is the same for gaining a follow in Twitter and joining Facebook.  These are very different actions in terms of effort required, and the overlap has a different meaning in each situation.  This suggests that the overlap feature may be important in a wide variety of social networks and its impact may be very similar.

We next perform a statistical analysis of the follow probability for this
experiment.
The features of our regression model are overlap and the log transformed friend and follower counts of the target.  We exclude the agent's friend and follower count because
or previous analysis found them not to be significant.   The resulting regression coefficients are shown in Table \ref{table: overlapreg}.  As can be seen, the most significant features are overlap and target follower count, while target friend count is significant at a 5\% level.  Increased overlap and target friend count increase the conversion rate, while increased target follower count decreases the conversion rate.

\begin{table} 
	\centering
	\caption{Regression coefficients for the follow probability. 
		The features are the log transformed friend and follower count of the target and the overlap.   	The significance codes are 
		$^{***}: 0.001$, $^{**}: 0.01$, and $^*: 0.05$.} \label{table: overlapreg}
	\begin{tabular}{|l||c||c|}
		\hline
		Feature & Coefficient & p-value\\ \hline 
		Intercept                          & -2.49*** & $4.10\times 10^{-8}$ \\ \hline
		Overlap                            &  0.28*** & $6.57\times 10^{-12}$  \\ \hline
		$\log_{10}$(Target Friend Count)   &  0.45* &  $0.05$ \\ \hline
		$\log_{10}$(Target Follower Count) & -0.63*** & $1.77\times 10^{-4}$  \\ \hline
	\end{tabular}
\end{table}

\section{Optimal Policies for the Follow-back Problem}\label{sec:theory}
We now study optimal policies for the follow-back problem.
We start by making assumptions on the structure of the follow probability.
We then derive results for the optimal policy and expected follows
on directed acyclic graphs (DAGs).
This analysis will form the foundation for our general solution to the follow-back 
problem in Section \ref{sec:ip}.

\subsection{Assumptions on Follow Probability Model}\label{sec:finalmodel}
To solve the follow-back problem we must make some assumptions on the follow probabilities. In our general model, the follow probability depended upon vertex features, graph state, and interaction type. We assume the agent uses a single type of interaction, so this is not a variable we need to optimize.  Based on our results in Section \ref{sec:Interaction} this interaction should be following and retweeting.   Our  analysis in Section \ref{sec:follow_count} suggests that the vertex features we should  consider are the friend and follower count of the target.  For a target $v$ we will call this set of features $F_v$.  Also, as our analysis in Section \ref{sec:overlap} showed, the graph feature we need to include is the overlap between the agent and target.  We define the overlap between the agent and target $v$  as $\phi_{v}$. 

With these assumptions we can write the follow probability of a target $v$ in response to an interaction by the agent as $p(\phi_v,F_v)$.  We further assume that the follow probability has the product form $p(\phi_{v},F_v)=f(\phi_{v})g(F_{v})$.  We define $g(F_v)$ as the \emph{susceptibility} of the vertex.  It is the follow probability if there is no overlap with the agent.  The product form assumption can be viewed as an approximation to the logistic form of the follow probability given by equation \ref{eq:logistic} when the probability value is small.  Assume the coefficient of the overlap is $\beta_\phi$ and the coefficient vector of the vertex feature set is $\beta_F$.  From our logistic regression model, the follow probability is given by
\begin{align} 
p(\phi_v,F_v) & = \frac{e^{\beta_\phi\phi_v+\beta_F^TF_v}}{1+e^{\beta_1\phi_v+\beta_F^TF_v}}\nonumber\\
&\approx e^{\beta_\phi\phi_v}e^{\beta_F^TF_v}\\
&\triangleq f(\phi_v)g(F_v).\label{eq:susceptibility} 
\end{align}
The above expression is a good approximation when the term in the exponent is much less than zero, which is a valid assumption when follow probabilities are only a few percent.  We use this approximation because it allows us to obtain simple analytic solutions for the follow-back problem on different graph topologies.

We assume that the follow probabilities are monotonically increasing in the value of the overlap, which roughly aligns with the trend we saw in Figure \ref{fig:overlap_plot} and the results of our regression analysis. Furthermore, recall that in Section \ref{sec:def}, we assumed that the agent's interaction with one user must conclude before it begins interacting with another user. In other words, the agent cannot begin interacting with one user, start communicating with another user, and switch back to interacting with the former. Since we have also fixed the interaction type, the follow-back problem reduces to choosing a sequence of vertices  with whom the agent interacts.  We now look at the structure of an  optimal sequence for a particular class of graph.

\subsection{Optimal Policy on a Directed Acyclic Graph (DAG) }
\begin{figure} [t]
	\centering
	\includegraphics[scale = 1]{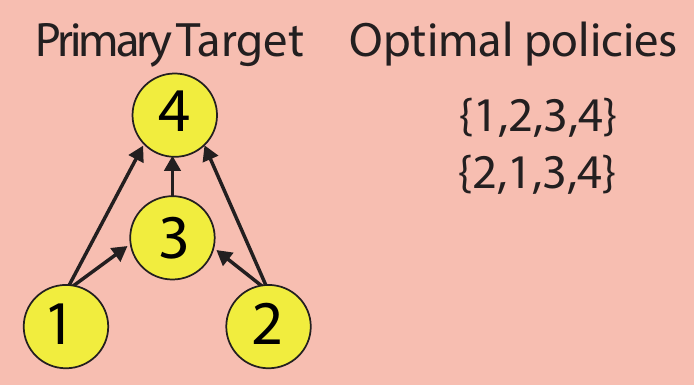}
	\caption{Illustration of two optimal linear extension policies on a DAG where all vertices are in the target set.  The optimal policies have the agent interact with a vertex only after it has interacted with its parents. } 
	\label{fig:dag} 
\end{figure}

On DAGs the analysis of the follow-back problem is tractable
because the graph contains no cycles. Since the follow probabilities are increasing
in overlap, the intuitive solution here would be to maximize the overlap for each interaction by following a vertex after following its parents.  To formalize this, recall that a DAG can also be viewed as a partially ordered set (poset) on the vertices, with the partial order requiring that a vertex must come after its parents.  Any sequence of vertices which respects this partial order is known as a linear extension.  We consider the situation where every vertex in the DAG is in the target set.  Under this condition,  we have the following result, which is proved in Appendix \ref{app:thmdag}.
\begin{theorem}\label{thm:dag}
	Assume that the follow-probability is monotonically increasing in the overlap.
	Let $G = (V,E)$ be a  directed acyclic graph and let all vertices in $V$ be in the target set.  
	Let $\pi^*$ be the optimal policy for the follow-back problem on $G$.  Then $\pi^*$ is a linear extension of $G$.
\end{theorem}
%
What is interesting about this result is that \emph{any} linear extension is an optimal policy.  This is because all linear extensions result in the same follow probabilities.  Intuitively, this is true because the follow probabilities depend only upon features of the target and the overlap.  Under our assumptions on the follow probabilities, interchanging the order of the vertices in the policy in a way that respects the partial order does not affect any of these features.  We illustrate linear extension policies for a DAG in Figure \ref{fig:dag}.   

\subsection{Optimal Follow Probability on a DAG}\label{sec:dag_follows}
We know that the optimal policy on a DAG is a linear extension.  A natural question to ask is what is the follow probability of a vertex under an optimal policy on a DAG?  Here we provide a closed form expression for this follow probability.   Recall that the follow probability for a target vertex $v$ with features set $F_v$ and overlap $\phi_v$ is given by $p(\phi_v,F_v) = f(\phi_v)g(F_v)$ as outline in Section \ref{sec:finalmodel}.  To obtain a simple closed form expression, we assume that $f(\phi_v)$ is monotonically increasing and linear, i.e. $f(\phi_v) = 1+\beta\phi_v$ for some $\beta>0$.  This is a good approximation for small overlap values because $\beta$ is small, as seen
from Table \ref{table: overlapreg}.      We let $\mathcal P_l(v,G)$ be the set of cycle free directed paths of length $l$ in a graph $G$ which terminate on vertex $v$.  For instance, if $v$ is a root vertex of a DAG $G$, then $\mathcal P_0(v,G)=\curly{v}$, and if $u$ is a child of $v$, then $\mathcal P_0(u,G)=\curly{u}$ and $\mathcal P_1(u,G)=\curly{(v,u)}$.  We have the following result for the follow probability under a linear extension policy on a DAG, which is proved in Appendix \ref{app:thmdag1}.
\begin{theorem}\label{thm:dag1}
	Let $G=(V,E)$ be a DAG with $N$ vertices.  Let the follow probability for a  vertex $v$ with features set $F_v$ and overlap $\phi_v$ be given by $p(\phi_v,F_v) = g(F_v)(1+\beta\phi_v)$ for some $\beta>0$.  Let $p_v$ be the follow probability of a vertex $v\in V$ under a linear extension policy.  Then 
	\begin{align}
	p_v & =  \sum_{l=0}^{N-1}\beta^l\sum_{P\in\mathcal P_l(v,G)}w_P \label{eq:dag1}
	\end{align}
	where for a path $P$ we define 
	\begin{align}
	w_P =\prod_{u\in P}g(F_u).\label{eq:weight}
	\end{align}
\end{theorem}
This expression weighs each possible  path $P$ of length $l$ by $\beta^lw_P$.  The follow probability is obtained by summing over all
possible paths in the DAG which terminate on the vertex.  
If we look at each term corresponding to a path of length $l$ in the expression for $p_v$ we can gain a better understanding of
how the overlap plays a roll in the optimal follow probability.  For each value $l$ we define
\begin{align}
q^{l}_v = \beta^l\sum_{P\in\mathcal P_l(v,G)}w_P \label{eq:dagl}
\end{align}
as the contribution to  $p_v$ coming from a specific value of $l$ in the sum in equation \eqref{eq:dag1}.  With this notation,
we can rewrite equation \eqref{eq:dag1} as 
\begin{align}
p_v = \sum_{l=0}^{N-1}q^l_v  .\label{eq:dag2}
\end{align}
For $l=0$ we have
\[q^0_v = g(F_v).\]
This is just the follow probability with no overlap. It can also be viewed as the contribution of zero hop paths to $p_v$. For $l=1$ we have
\[ q^1_v = \beta\sum_{u:(u,v)\in E}g(F_u)g(F_v),\]
which is simply the sum over all the edges in the graph that terminate on $v$, with weights given by the product of the susceptibility of
their end vertices.  This can also be viewed as the one hop path contribution
to $p_v$.   For $l=2$, we have
\[ q^2_v = \beta^2\sum_{u,w:(w,u),(u,v)\in E}g(F_w)g(F_u)g(F_v),\]
which is the contribution of two hop paths that terminate on $v$.
Continuing this way we obtain contributions
from higher order paths to $p_v$. The higher order terms
decrease exponentially in $\beta$, but include sums whose number
of terms depend upon the number of paths of a given length
and the vertex susceptibilities.  We will generally operate in
a regime where the susceptibilities are low and $\beta$ is much less
than one.  This will allow us to accurately approximate $p_v$ with either
a first or second order expansion.  Using these low order expansions
for the optimal follow probability will allow for tractable solutions
to the follow-back problem on arbitrary graphs, as we will see in Section \ref{sec:ip}.

\section{Integer Programming Formulation of the Follow-back Problem} \label{sec:ip}
We now consider solving the follow-back problem on an arbitrary graph.  We saw that for DAGs the optimal policy was a linear extension.  We also showed how to calculate the  follow probability for a vertex under a linear extension policy.  In this section we show how to extend these results to solve the follow-back problem on an arbitrary graph.  Our approach utilizes approximations for the follow probability in a clever integer programming formulation.

\subsection{Expected Follows on Graphs with Cycles}
The expression for the  optimal follow probability in equation \eqref{eq:dag1} assumes the underlying graph is a DAG.  However, this expression is still useful for graphs with cycles.  Consider an interaction policy (sequence of vertices) for an arbitrary graph.  The policy imposes an ordering on the vertices.  Some edges in the graph do not respect this ordering.  To see the role these edges play,  consider two vertices $u$ and $v$ in a graph where there is an edge from $u$ to $v$ and consider a policy where $v$ comes before $u$.  If the agent gains a follow from $u$, this increases its overlap with with $v$ because of the $(u,v)$ edge.  However, since the agent
already interacted with $v$ before $u$ in the policy, this overlap has no effect
on the follow probability of $v$.     In fact, any edge that
does not respect the policy will not impact any vertice's follow probability.  Therefore, once a policy is chosen, it creates a DAG on the graph and the follow probabilities
of vertices under the policy are given by equation \eqref{eq:dag1} with the constraint
that one only considers edges in the corresponding DAG.  Therefore, to solve the follow-back problem, we need to select the best DAG.

We want to choose a DAG such that the expected number of targets following the agent, which equals the sum of their follow probabilities, is maximized. To formalize this, consider a graph $G = (V,\mathcal E)$ with $|V|=N$. Let $\mathcal D(G)$ be the set of DAGs of $G$ and let $T\subseteq V$ be the set of target vertices.  Then using equation \eqref{eq:dag1}, the followback problem can be written as 
  \begin{align}
 \max_{D\in\mathcal D(G)} \sum_{v \in T}  \sum_{l=0}^{N-1}\beta^l\sum_{P\in\mathcal P_l(v,D)}w_P.\label{eq:followback_obj_1}
 \end{align}
 Here the sum is only taken over vertices in $T$, forcing the policy
 to focus on the target set.
We next look how to formulate this problem as an integer program.

\subsection{Integer Programming Formulation}
Solving the follow-back problem requires one to optimize over the set of DAGs of a graph.  We do this by using variables for all of the edges in the graph.  
We assume we are given an arbitrary directed graph $G = (V,\mathcal E)$. $V$ is the set of vertices in the graph and $|V| = N$. We let $E$ represent the adjacency matrix of the graph, such that $E_{uv} = 1$ if the directed edge $(u,v)\in\mathcal E$. We are given a subset of targets $T \subseteq V$ from which the agent seeks to gain followers. For each vertex $v\in V$ we denote its susceptibility as $g_v = g(F_v)$.  

To begin our integer programming formulation, we define the binary  variables
$x_v$ for $v\in V$ and $y_{uv}$ for $u,v\in V$.
The $x$ variables indicate which vertices the agent interacts with and the $y$
variables indicate which edges are in the DAG.  These variables determine
the DAG, and hence the interaction policy for the agent.

The first constraint in the follow-back problem concerns the agents interactions.
In practice, an agent cannot interact with every vertex in the graph.  Doing this
would make the agent appear like an automated bot and may reduce its effectiveness.
Therefore, we set an upper limit $m$ for the number of interactions for the agent.  This results in the constraint $\sum_{v \in V} x_v\leq m$.

Each $y$ variable can only be equal to one if it corresponds to an edge in the graph.  This can be done with the constraint $y_{uv}\leq E_{uv}$.  In addition, if the agent interacts with a pair of vertices which form an edge, then this edge must be in the DAG.  This is enforced with the  constraints:
\begin{align}
x_i - y_{i,j} \geq 0, &~~ \forall i, j\in V\label{eq:xy1}\\
x_j - y_{i,j} \geq 0, &~~ \forall i, j\in V\label{eq:xy2}. 
\end{align}
These constraints are sufficient because as we will see, to maximize the objective
function, the integer program will try to make all $y$ variables equal to one.  The constraints only allow a $y$ variable to be one if the corresponding $x$ variables
are also one.

Finally, the $y$ variables must correspond to edges which form a DAG.  This means there can be no cycles.  We define the set of all cycles in $G$ as $\mathcal C$.  Then the no cycle constraints can be written as 
\begin{align}
\sum_{(u,v) \in C} y_{uv} \leq |C| -1, ~~ \forall C \in \mathcal{C}.\label{eq:cycle}
\end{align}
The objective function for the follow-back problem is simply a weighted
sum of all paths in the chosen DAG that terminate on a target vertex.  With our variables, this can be written as
\begin{align}
\sum_{t \in T} g_tx_t + \sum_{t \in T} \sum_{l=1}^{N-1}\beta^l \sum_{P \in \mathcal P_l(t,G)}  \prod_{u \in P}g_u\prod_{(v,w) \in P}y_{vw}.\label{eq:obj_nonlinear}
\end{align}
Written this way we see that the objective function is non-linear and involves
products of the $y$ variables.  We will next look at linear approximations
to this function that allow for simple and fast solutions.  

\subsection{Linear Integer Programming Formulation}
The objective function is a sum over paths in the chosen DAG which terminate on a target.  The sum can be taken over paths of different lengths.  This allows us to define different integer programming formulations by the longest path length included. The zeroth order formulation considers only paths of length zero.  The objective function is
\begin{align}
\sum_{t \in T} g_tx_t.\label{eq:obj0}
\end{align}
The optimal solution to this problem is to select the targets with the highest susceptibilities, which makes intuitive sense.  However, the zeroth order formulation totally ignores the effect of overlap.  To include this, we can use a first order formulation with objective function
\begin{align}
\sum_{t \in T} g_tx_t + \beta\sum_{t \in T}\sum_{u\in V}g_ug_t y_{ut}.\label{eq:obj1}
\end{align}
Here the overlap is included for only the immediate parents of the targets and has a contribution of order $\beta$. However, any overlap between non-target vertices are not included.  

We can define a second order formulation to account for overlap between non-target vertices within two hops of a target as follows.  Define binary variables $z_{uvw}$ for $u,v,w\in V$.  We want these variables to be one
only if edges $(u,v)$ and $(v,w)$ are in the DAG.  This can be done with the constraints
\begin{align}
y_{uv}-z_{uvw}\geq 0, ~~ \forall u,v,w\in V \\
y_{vw}-z_{uvw}\geq 0,~~ \forall u,v,w\in V.
\end{align}
The second order objective function is
\begin{align}
\sum_{t \in T} g_tx_t + \beta\sum_{t \in T}\sum_{u\in V}g_ug_t y_{ut}
+\beta^2\sum_{t \in T}\sum_{u\in V}\sum_{v\in V}g_ug_vg_tz_{uvt}.\label{eq:obj2}
\end{align}
This function includes terms of order $\beta^2$ which come from the vertices two hops from targets.  To maximize this objective, all the $z$ variables will be set to one.  The constraints
ensure that this can only happen if the corresponding edges are included in the DAG.

We could continue this process for longer paths, but the resulting formulations would become too large to solve.  If paths of length $l$ are included, the number of variables in the formulation would scale like $|V|^l$.  This is reasonable for $l\leq 2$, but beyond that it can become intractable.  Also, it is not clear  if higher order formulations perform that much better than the first and second order formulations on real network topologies.  We will study this more closely in Section \ref{sec:results}.

One issue with our formulations is that the linear approximation for the follow probabilities of the targets can exceed one if the overlap is too high.  Recall that we obtained this approximation by performing a Taylor approximation to the logistic function in equation \eqref{eq:logistic}.  One way to fix this problem is to include a constraint that limits the value of the follow probabilities of the targets.  For the first order formulation the constraints are
\begin{align}
 g_tx_t + \beta\sum_{t \in T}\sum_{u\in V}g_ug_t y_{ut}\leq 1, ~~\forall t\in T,
\end{align}
and for the second order formulation the constraints are
\begin{align}
 g_tx_t + \beta\sum_{t \in T}\sum_{u\in V}g_ug_t y_{ut}
+\beta^2\sum_{t \in T}\sum_{u\in V}\sum_{v\in V}g_ug_vg_tz_{uvt}\leq 1, ~~\forall t\in T.
\end{align}
Without these constraints, the integer program may target a large number of parents of a target, mistakenly thinking this will linearly increase the follow probability.
These constraints capture the diminishing returns behavior of the follow probability
with respect to the overlap. They force the integer program to avoid targeting too many
parents of single target.  

To summarize, we have three different  linear formulations for the follow-back problem.  Each provides a closer approximation to the non-linear objective function in equation \eqref{eq:obj_nonlinear}, but at the cost of larger formulations.  The simplest is the zeroth order formulation, which ignores all edges, and is given by
\begin{equation}\label{eq:IP0}
\begin{aligned}
& \max_{\mathbf{x}} \sum_{t \in T} g_tx_t  \\
\text{Subject to} &\\
& \sum_{u \in V} x_u \leq m\\
& x_u \in \curly{0,1} &\forall u\in V.
\end{aligned}
\end{equation}
The first order formulation, which looks only at one hop paths terminating
on target vertices, is given by
\begin{equation}\label{eq:IP1}
\begin{aligned}
& \max_{\mathbf{x},\mathbf{y}} \sum_{t \in T} g_tx_t + \beta\sum_{t \in T}\sum_{u\in V}g_ug_t y_{ut}\\
\text{Subject to} &\\
& \sum_{u \in V} x_u \leq m\\
& y_{uv} \leq E_{uv}, & \forall u,v \in V\\
& \sum_{(u,v) \in C} y_{uv} \leq |C| -1, & \forall C \in \mathcal{C}\\
& x_u - y_{uv} \geq 0, & \forall u,v \in V\\ 
& x_v - y_{uv} \geq 0, & \forall u,v \in V\\ 
& g_tx_t + \beta\sum_{t \in T}\sum_{u\in V}g_ug_t y_{ut}\leq 1, &\forall t\in T\\
& x_u \in \curly{0,1}, &\forall u\in V\\
& y_{uv} \in \curly{0,1},&\forall u,v \in V. 
\end{aligned}
\end{equation}
The second order formulation, which looks at one and two hop paths terminating
on target vertices, is given by
\begin{equation}\label{eq:IP2}
\begin{aligned}
& \max_{\mathbf{x},\mathbf{y},\mathbf z} \sum_{t \in T} g_tx_t + \beta\sum_{t \in T}\sum_{u\in V}g_ug_t y_{ut}
+\beta^2\sum_{t \in T}\sum_{u\in V}\sum_{v\in V}g_vg_ug_tz_{uvt}\\
\text{Subject to} &\\
& \sum_{u \in V} x_u \leq m\\
& y_{uv} \leq E_{uv}, & \forall u,v \in V\\
& \sum_{(u,v) \in C} y_{uv} \leq |C| -1, & \forall C \in \mathcal{C}\\
& x_u - y_{uv} \geq 0, & \forall u,v \in V\\ 
& x_v - y_{uv} \geq 0, & \forall u,v \in V\\ 
&z_{uvw}-y_{uv}\leq 0, & \forall u,v,w\in V \\
&z_{uvw}-y_{vw}\leq 0,& \forall u,v,w\in V\\
& g_tx_t + \beta\sum_{t \in T}\sum_{u\in V}g_ug_t y_{ut}
+\beta^2\sum_{t \in T}\sum_{u\in V}\sum_{v\in V}g_ug_vg_tz_{uvt}\leq 1, &\forall t\in T\\
& x_u \in \curly{0,1}, &\forall u\in V\\
& y_{uv} \in \curly{0,1},&\forall u,v \in V\\ 
& z_{uvw} \in \curly{0,1},&\forall u,v,w \in V. 
\end{aligned}
\end{equation}


\section{Results}\label{sec:results}
Here we present results for our different follow-back problem formulations based on
simulations on two real graph topologies.  
The first graph has one target, U.S. President
Donald Trump, his friends, and his friends friends. 
The other graph consists of 11 targets and all of their friends.   For each graph we compare our follow-back policies to other policies.  Throughout this section we will use the parameter values from Table \ref{table: overlapreg}.  We set $\beta = 0.28$ for the overlap.  For a vertex $v$, let us define its friends and follower counts as $d^{in}_v$ and $d^{out}_v$, respectively.  Then, applying equation \eqref{eq:susceptibility}  and the values in Table \ref{table: overlapreg}, the susceptibility for $v$  is given by
\begin{align}
	g_v & = \exp\paranth{-2.49+0.45\log_{10}(d^{in}_v+1)-0.63\log_{10}(d^{out}_v+1)}.
\end{align}
These susceptibility values and our value for $\beta$ are used in our formulations to determine the policies.

To solve our integer programming formulations we use the Julia Math Programming (JuMP) package in Julia \citep{JuMP} with Gurobi as the solver \citep{gurobi}.  
One concern with our formulations is the potentially exponential number of constraints related to eliminating cycles in the graph. To handle this,  we rely upon  lazy constraint generation \citep{lazy_constraint}.  The basic idea is to iteratively add in
cycle constraints until one obtains a feasible solution (a DAG). We utilize the lazy constraint generation function in JuMP to solve our formulations \citep{DunningHuchetteLubin2017}.

\subsection{Non-Followback Policies}
We compare the follow-back policies with a random policy we refer to as \emph{random append} which selects a random permutation of  members of the friends graph and then appends a random permutation of the target users.  This policy selects the targets' friends before the targets, but does not use any other graph information to optimize the vertex sequence.
We also compare the follow-back policies to network centrality based polices.  These policies rank the vertices based on the value of the network centrality.    To be precise, for a network centrality policy, we first calculate the network centrality of each vertex in the graph, including the targets.  Then, we rank the vertices by either increasing or decreasing centrality, while being sure to append the targets to the end of the sequence.  This then forms the centrality based policy.  We use the susceptibilities as the
vertex weights when calculating the centralities.  We looked at many centrality measures, but found that the best performing one was eigenvector centrality, which we focus on here.

\subsection{Simulation Methodology}\label{sec:simulation}
We evaluate the performance of a policy as follows.
We simulate an interaction between the agent and each vertex of the policy in sequence.   The result
of this interaction is given by a Bernoulli random variable that is one with probability
given by the logistic function in equation \eqref{eq:logistic} with the coefficient values from  Table \ref{table: overlapreg}.  The features of the interaction are the friend count, follower count, and overlap as a result of previous interactions in the policy.  We then update the overlap of all vertices in the graph as a result of the interaction.  In particular, if this vertex
follows back, then all of its children in the graph increase their overlap by one.  
We repeat this process for each vertex in the policy.  When we reach the end of the policy, we measure which of the  targets  follow the agent.   We simulate each policy 10,000 times to obtain the expected target follows of the policy.

\subsection{Single Target: realDonaldTrump}
Here we study the follow-back problem for a single target:  U.S. President Donald Trump who has the Twitter screen name realDonaldTrump.  There are 45 immediate friends of realDonaldTrump and 1,000 two hop friends.  Detailed properties of this graph are shown in
Table \ref{table:target_graph}.  If we use the first order formulation, then the agent will not interact with anyone who is not a neighbor of the target.  This is because the objective function only includes edges which contain the target.  However, the second order formulation will take into account edges that contain the target's friends.  Therefore, using the second order formulation allows us to obtain the benefits of using a deeper graph when trying to connect with the target.


\begin{table}\label{table:target_graph}
	\centering
	\caption{Graph properties for realDonaldTrump's two hop friends graph and the multiple targets one hop friends graph.}
	\begin{tabular}{|c|c|c|}
		\hline
		Graph                          & realDonaldTrump two hop& Multiple targets one hop\\
		                               & friends graph       & friends graph\\ \hline\hline
		Number of targets              & 1     &  11\\ \hline
		Number of vertices             & 1,046 &   2,806 \\ \hline
		Number of edges                & 14,870 &   83,201\\ \hline
		\end{tabular}
\end{table}

\begin{figure} 
	\centering
	\includegraphics[scale = .5]{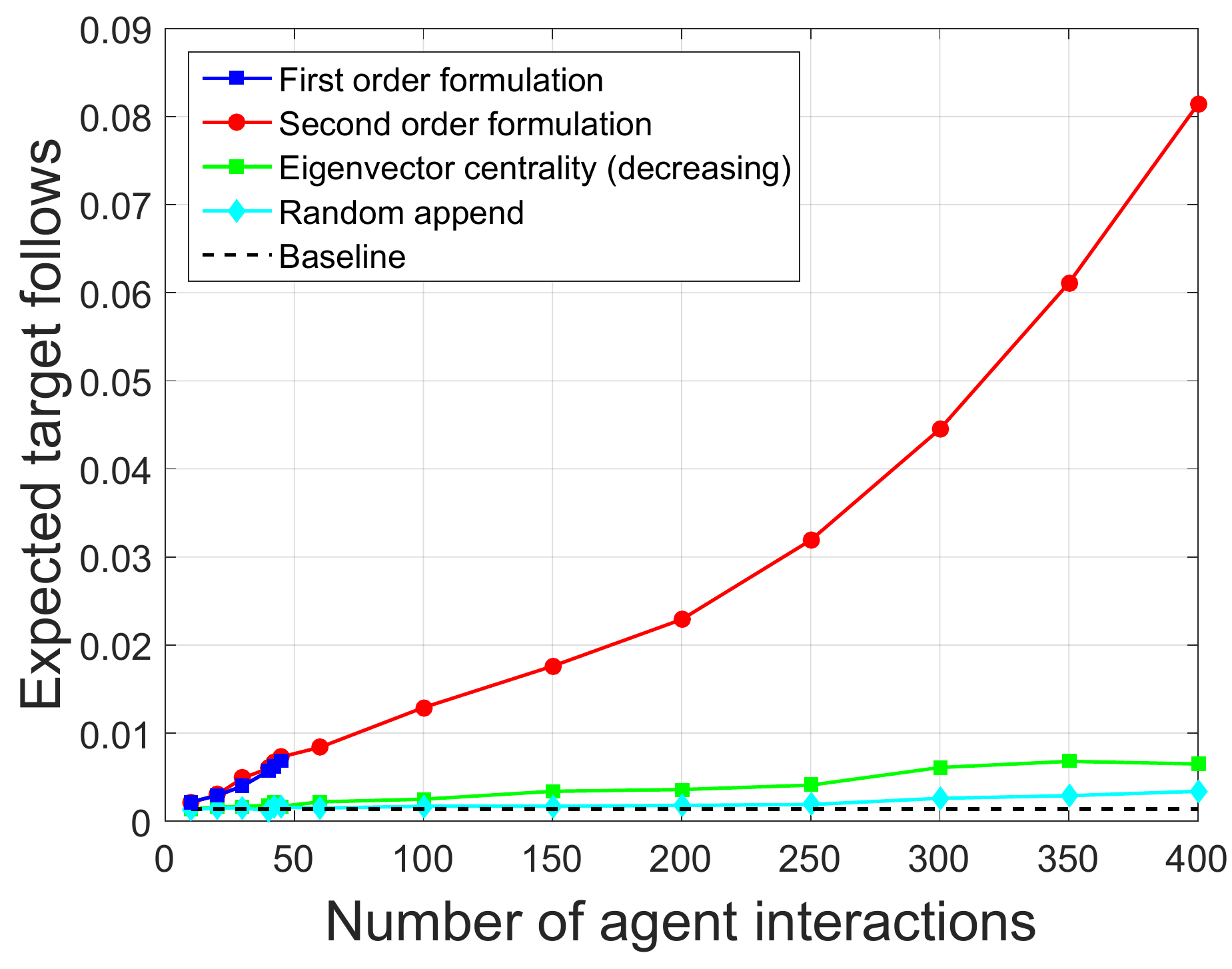}
	\caption{Plot of expected target follows versus number of agent interactions using different policies for  realDonaldTrump's two hop friends graph.  } 
	\label{fig:trump_m} 
\end{figure}

  We test different values for the number of agent interactions in the follow-back policies.  The resulting expected follows of the different policies are shown in Figure \ref{fig:trump_m}.  The susceptibility of realDonaldTrump is $0.0014$.  This is the baseline
  follow probability.  As can be seen, our policies exceed this value
  by a significant amount.  The best performing policies are the first and second order integer programming formulations.  Random and eigenvector centrality polices perform
nearly an order of magnitude worse.  With 400 interactions, the follow probability of the target is 0.08, while for the random and centrality policies the follow probability is less than 0.01.

The first order formulation policy is only calculated up to 45 interactions.  This is because this is the number of friends the target has.  The second order formulation policy
is nearly equal to the first order in this regime.  This suggests that the first order formulation may be a good approximation to the second order formulation in realistic instances of the follow-back problem.  

\begin{figure} 
	\centering
	\includegraphics[scale = .5]{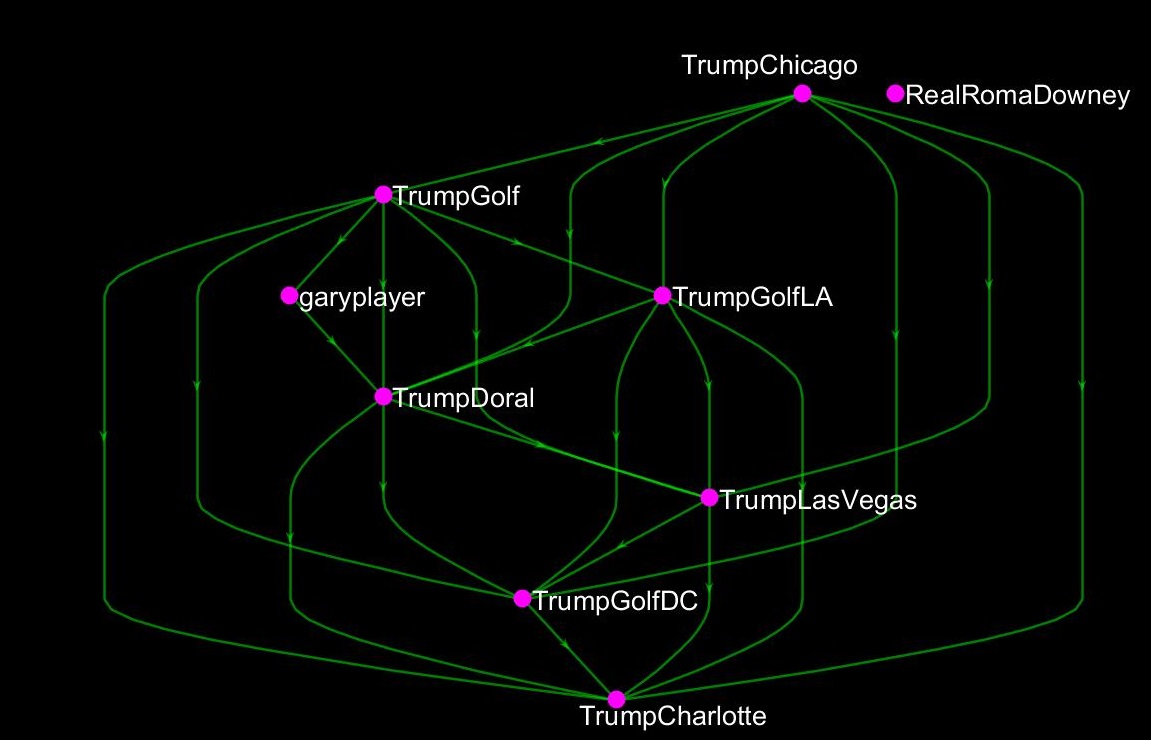}
	\caption{The DAG for the second order formulation interaction policy with 10 interactions on realDonaldTrump's two hop friends graph.    } 
	\label{fig:dag10} 
\end{figure}

\begin{figure} 
	\centering
	\includegraphics[scale = .5]{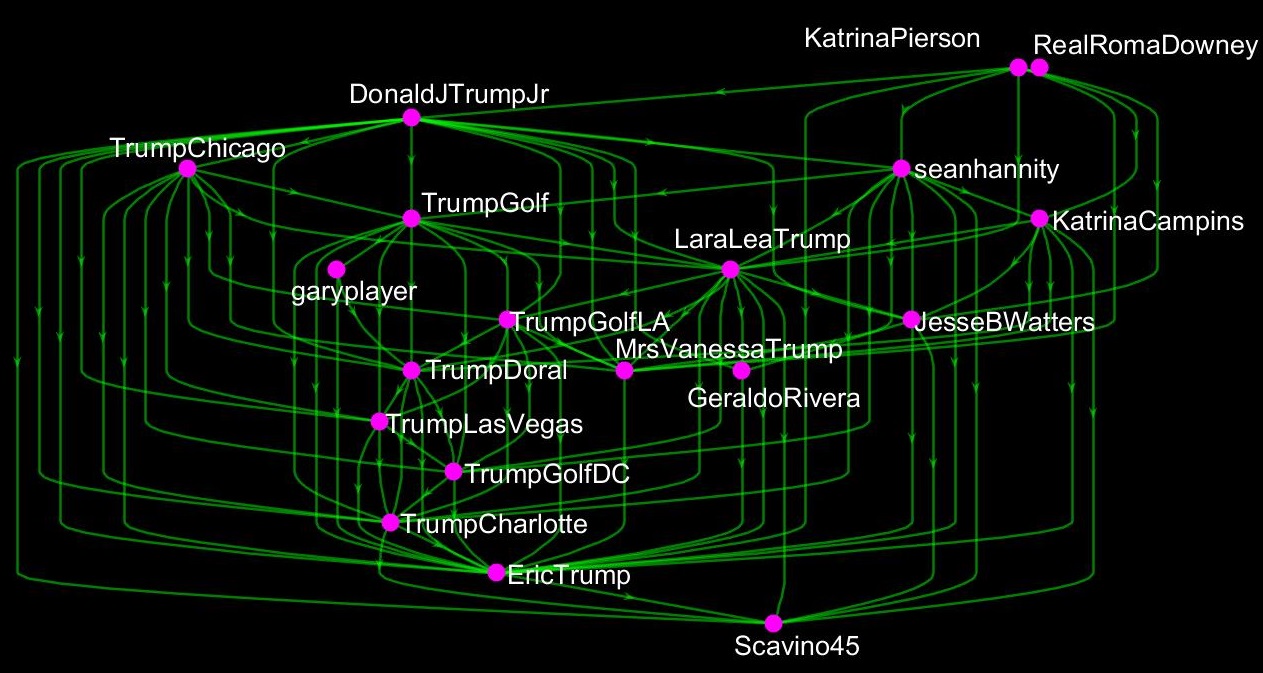}
	\caption{The DAG for the second order formulation interaction policy with 20 interactions on  realDonaldTrump's two hop friends graph.    } 
	\label{fig:dag20} 
\end{figure}

We show in Figures \ref{fig:dag10} and \ref{fig:dag20} the DAGs for the second order formulation policy
with 10 and 20 interactions, respectively.  We do not include the target in the figure in order to reduce
the number of edges and improve readability of the vertex names.  

In the 10 vertex policy the majority of vertices have high susceptibilities.  This
is because they generally  have few followers (few in this case means between 80,000 and 200,000).  Many of the vertices are Twitter accounts of properties owned by the Trump Organization and are not very influential celebrities.  The two exceptions are Gary Player (screen name @garyplayer) and Roma Downey (screen name @RealRomaDowney).  Both of these individuals have over 2,000 friends, which gives them a high susceptibility, despite having several hundred thousand followers.  
 
The 20 vertex policy contains many famous or influential individuals.
The first vertex in the policy is Katrina Pierson (screen name @KatrinaPierson), who was the national spokesperson for the Donald Trump 2016 presidential campaign.  She has similar friend and follower counts as Gary Player, but has no friends among the 20 vertices in the policy, and hence no overlap to gain from them.    The last vertex in the policy is Dan Scavino (screen name @Scavino45) who is the White House Director of Social Media and has seven friends among the vertices in the DAG.  Generally speaking, vertices with more friends come later in the policy in order to build overlap.  Many network centralities have the property that higher degree vertices have higher values.  This suggests that centrality based polices should be similar to our integer programming based policies.  However, this is not always the case.  For instance, Eric Trump (screen name @EricTrump) who is the son of Donald Trump, has 16 friends among the vertices in the DAG, but comes before Dan Scavino who has only 7.  We believe that the complex interaction of vertex susceptibilities and graph structure is why our integer programming policies differ from centrality based policies.

\subsection{Multiple Targets}
We next consider the follow-back problem with multiple targets.  We selected 11 Twitter users that span multiple fields.  The list of the targets and their basic properties are shown in Table \ref{table:targets}.  We solve the follow-back problem on the graph consisting of these targets and their friends.  This is in contrast to the single target graph where we also included two hop friends.  We did not consider the two hop friends for the multiple target graph because rate limitations imposed by the Twitter API prevented us from collecting all of the edges.  Because we do not have the two hop friends, we will only
calculate the first order formulation policy.  This means the only edges considered
are those between targets and between the targets and their friends.  Basic properties of the multiple target one hop friends graph are shown in Table \ref{table:target_graph}.  As can be seen, there are over 2,806 vertices in the graph and only 11 targets.  

\begin{figure} 
	\centering
	\includegraphics[scale = .5]{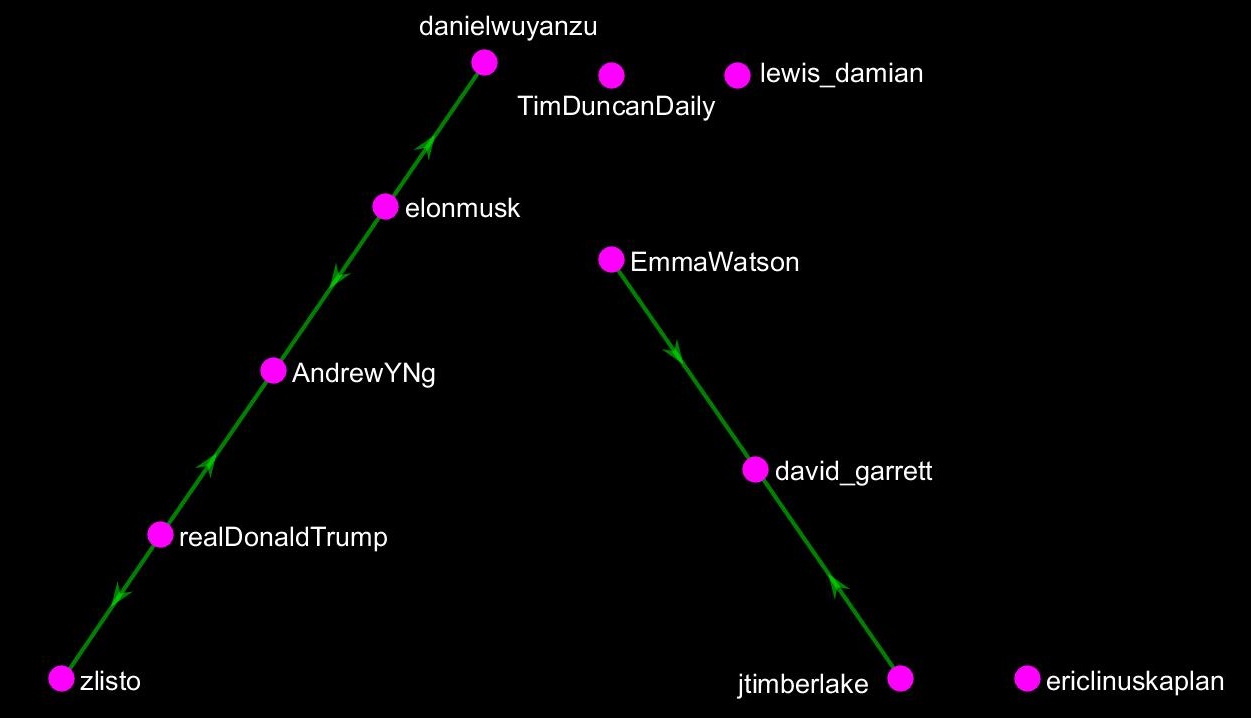}
	\caption{The subgraph induced by the targets in the multiple targets graph.    } 
	\label{fig:targets_graph} 
\end{figure}

We show the subgraph induced by the targets in Figure \ref{fig:targets_graph}. As can be seen, the targets form a DAG.  If we were to only interact with the targets, then the first few vertices in the optimal policy would include realDonaldTrump, elonmusk,
jtimberlake, and EmmaWatson  because they do not have any parents, and therefore the agent can
have no overlap with them.  However, these are influential figures
with low baseline follow probabilities, so it is unlikely the agent would
get any overlap with the other targets.  This is why we want to
include the friends of the targets, because many times these friends
have higher follow probabilities and there is a better chance of gaining
overlap with a target.  

\begin{table}\label{table:targets}
	\centering
	\caption{Basic features and information of the targets.  The targets are listed in order of decreasing baseline follow probability.}
	\begin{tabular}{|c|c|c|c|c|c|}
		\hline
		\textbf{Twitter } & \textbf{Name} &\textbf{ Background} & \textbf{Friend } & \textbf{Follower} &\textbf{Baseline} \\
		 \textbf{screen name} & & & \textbf{count} & \textbf{count} &\textbf{ probability}\\ \hline
		zlisto & Tauhid Zaman & Professor &496 & 957 & 0.0403 \\ \hline
		ericlinuskaplan &  Eric Kaplan & TV writer & 413 & 3129& 0.0268\\ \hline
		TimDuncanDaily & Tim Duncan & Basketball player & 305 & 46,900& 0.0132\\ \hline
		david\_garrett  & David Garrett & Violinist & 457 & 111,000& 0.0113\\ \hline
		  danielwuyanzu & Daniel Wu & Actor&  333 & 230,000& 0.0090\\ \hline
		AndrewYNg & Andrew Ng & Professor & 342 & 273,000 & 0.0087\\ \hline
		lewis\_damian & Damien Lewis & Actor & 63 & 177,000& 0.0067\\ \hline
		EmmaWatson & Emma Watson & Actress & 383 & 28.8 million& 0.0024 \\ \hline	
		elonmusk & Elon Musk & Entrepreneur & 48 & 20.5 million& 0.0019 \\ \hline	
		jtimberlake & Justin Timberlake & Musician & 262 & 65.5 million& 0.0018\\ \hline	
	    realDonaldTrump &  Donald Trump & U.S. President & 45 & 49.3 million & 0.0014\\ \hline
	\end{tabular}
\end{table}

\begin{figure} 
	\centering
	\includegraphics[scale = .5]{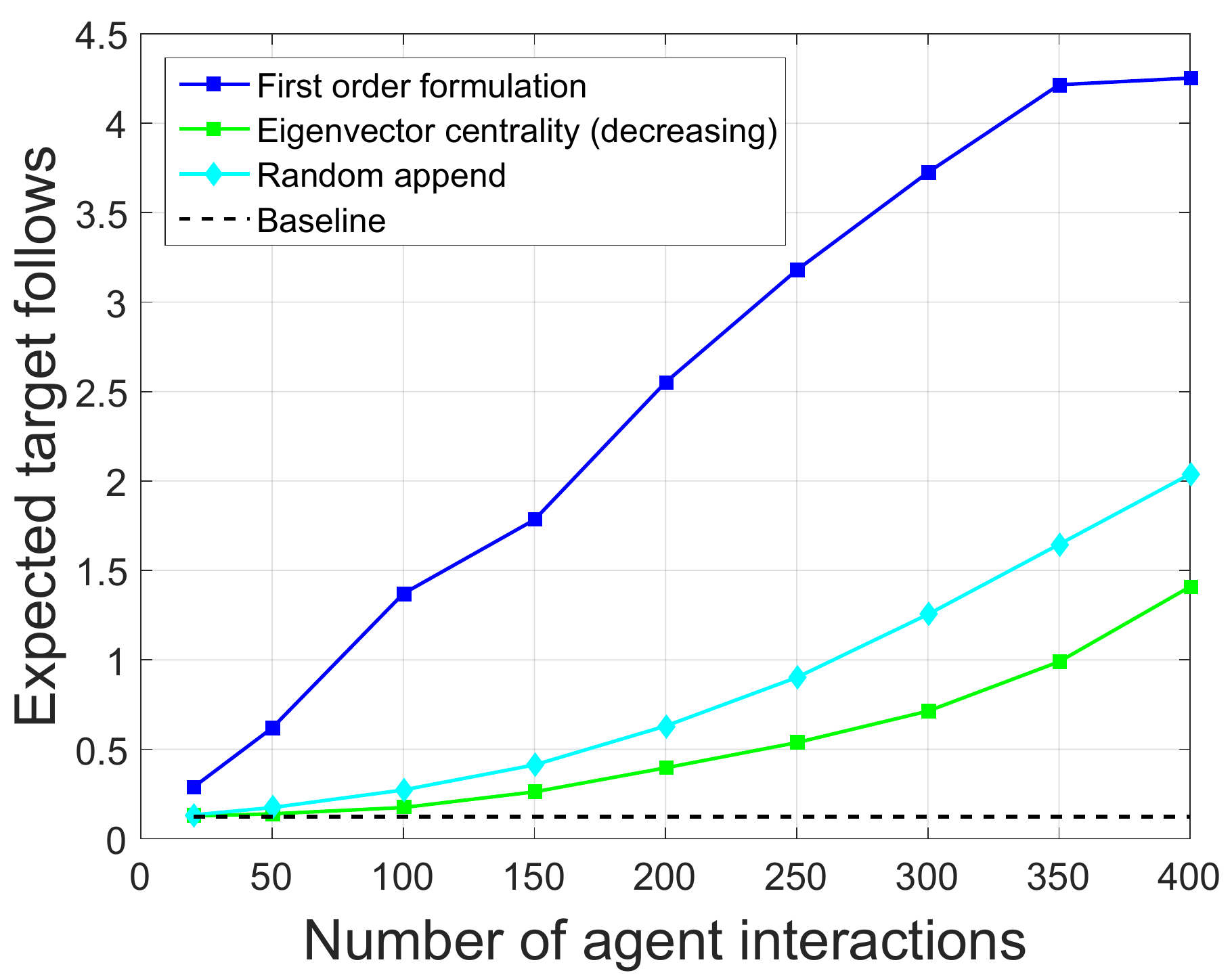}
	\caption{Plot of expected target follows versus number of agent interactions using different policies for the multiple targets one hop friends graph.    } 
	\label{fig:targets_m} 
\end{figure}

Figure \ref{fig:targets_m} shows how the different policies perform
on the multiple targets graph as a function of the number of agent interactions. 
As with the one target graph, the first order policy outperforms the centrality and random policy.
For 200 interactions, the first order policy is nearly an order of magnitude
better, resulting in 2.5 expected follows from the 11 targets.  To better see the impact of including the targets' friends, we show the expected follows 
for the baseline (sum of target follow probabilities with zero overlap), target subgraph (no
interactions with targets' friends), and the 200 interaction policies in Table \ref{table:target_policy}.  As can be seen, using an optimal policy
on the target subgraph does not change the expected follows very much at all.
This is because the targets with no parents have very low follow probabilities.
There is a very small chance that the agent will gain any overlap
with their children, and therefore they do not change their children's
follow probability very much.
When we include the friends of the targets, the expected follows
jumps by a factor of 20 over the optimal policy on the target subgraph.
This is due to the fact that these friends have higher follow probabilities,
so there is a better chance the agent gains an overlap with the targets.

\begin{table}\label{table:target_policy}
	\centering
	\caption{Expected follows of different policies and different subgraphs
		of the multiple targets one hop friends graph.}
	\begin{tabular}{|l|l|c|}
		\hline
		Policy & Graph & Expected follows\\ \hline
		Baseline & None & 0.123 \\ \hline
		Optimal&  Multiple targets subgraph & 0.124 \\ \hline
		First order, 200 agent interactions  & Multiple targets one hop friends graph &  2.552 \\ \hline
		\end{tabular}
\end{table}

\begin{figure} 
	\centering
	\includegraphics[scale = .9]{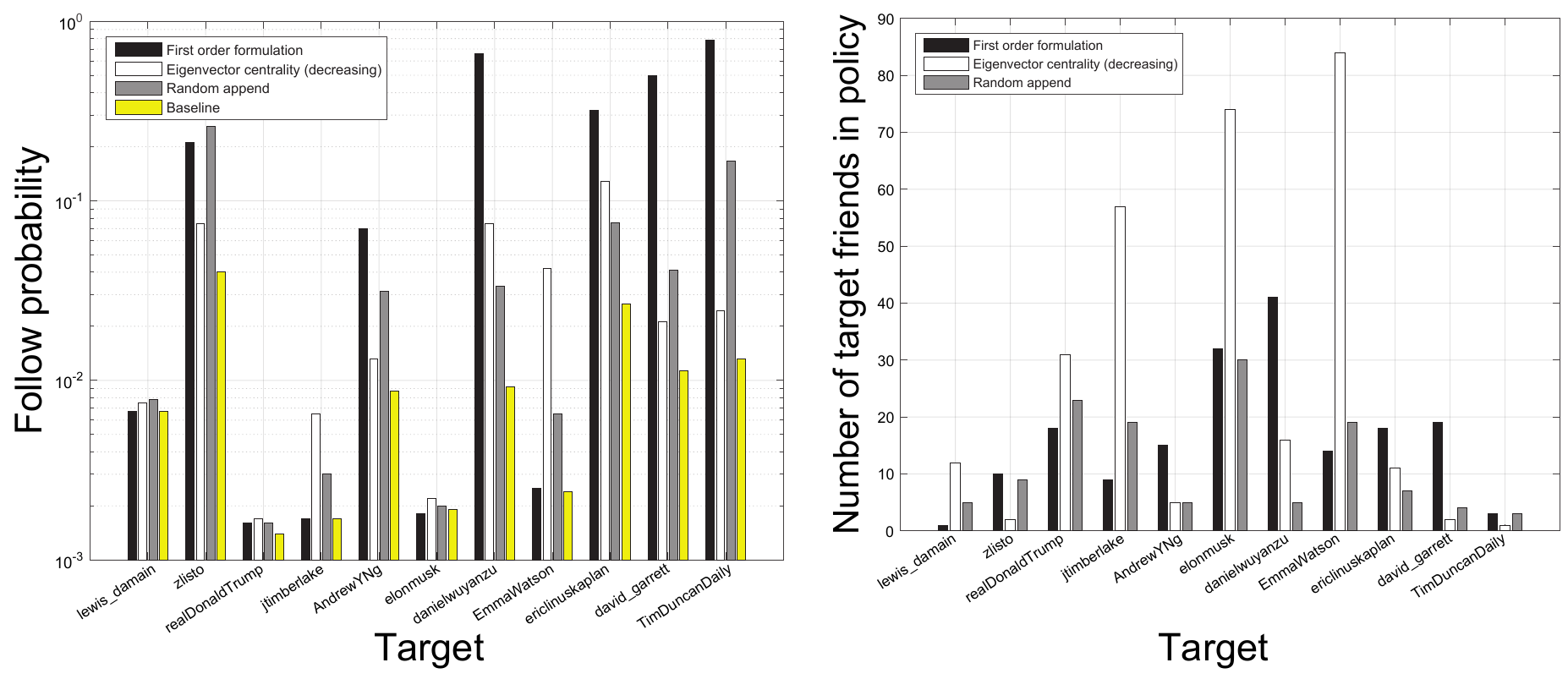}
	\caption{(left) Follow probability of each target under different policies with 200 agent interactions on the multiple targets one hop friends graph. (right) Number of friends of each target included under different policies with 200 agent interactions on the multiple targets one hop friends graph.    } 
	\label{fig:target_friend} 
\end{figure}

We next explore the 200 interaction first order policy in more detail.
Figure \ref{fig:target_friend} shows how the policy breaks down
across individual targets in terms of expected follow probability in the resulting
DAG and number of friends in the policy.  We first see that the policy is
focusing on friends of select targets, such as danielwuyanzu, AndrewYNg, and david\_garrett.  These individuals
have a high baseline follow probability, and so by targeting them, the agent obtains a significant increase in the total expected target follows.  This targeting can also be seen in the increase of their follow probabilities under the first order policy compared to the centrality and random policies.

An interesting observation is that the first order policy targets much fewer
friends than the centrality policy for many of the more famous targets such as emmawatson, elonmusk, and jtimberlake.  It is likely that their friends are also very influential, and so have a higher centrality value.  However, it is also likely that they have a low follow probability and will give no overlap with targets.  Therefore, even though these friends are highly central, they do not increase expected target follows.  The first order policy takes this fact into account.  In fact, the first order policy does nearly the opposite of the
centrality policy.  It looks for non-central friends, who are also likely to be
less influential and have higher follow probabilities.  These less central friends
act as vulnerabilities that allow us to penetrate the targets.

\section{Conclusion}\label{sec:conclusion}
We have proposed  the follow-back problem for penetrating a social network.    Empirical studies in Twitter identified features and interaction types that cause users to follow an agent.  We found that vulnerable users have many friends, few followers, and a high overlap with the agent.  This allowed us to identify vulnerabilities in a social network and operationalize network penetration.  Based on these empirical results, we developed a simple model for the follow probability.  We then used this model to derive optimal policies for the follow-back problem on DAGs.  We used the results for DAGs to develop an integer programming formulation for the follow-back problem on arbitrary graphs.  Simulations on real network topologies showed significant improvements in expected target follows of an agent using our integer programming policies.  

The follow-back problem focused on establishing a connection with a set of targets.  This is the first step in any social network based counter-measure.  Once a connection is established, one can begin implementing the counter-measure.  Generally, this involves
pumping information into the targets in an effort to either counter an influence
campaign or mitigate propaganda from extremist groups.  It is an important
area of future work to understand how to identify targets and pump information
into the targets to modify their opinions.  However, we emphasize that none of these counter-measures  is possible until a connection is formed between the targets and agents.

The work developed here on penetrating social networks has application in domains
beyond national security.  The crux of the follow-back problem is making a target
connect with an agent, in order to allow the agent to modify the target opinion.
This target can be an extremist or a voter.  It could also be a customer for
a product.  Our work here could immediately be applied in advertising to improve
targeted marketing campaigns.  Instead of using online advertisements, one could
use online social media accounts to spread information.  This may be more effective
than an advertisement if the target forms a bond of trust with the account.

We conclude by noting that despite the applicability across different domains,
 the ability to perform such
opinion manipulation over large populations
 should  be used judiciously. Manipulating the flow of information to target audiences using artificial social media accounts raises many ethical concerns and can have tremendous impact over the targets' understanding of world events and their subsequent actions.  While there is great potential for such social network counter-measures, there are also serious dangers.  These capabilities should be employed only when the situation truly warrants their use.  In national security applications, it is important that these types of capabilities are used only under supervision of leadership in the intelligence or military communities.

\ECSwitch

\ECHead{Supplementary Material and Proofs of Statements}

\section{Proof of Theorem \ref{thm:dag}}\label{app:thmdag}
Let the friends graph be $G=(V,E)$ which is a DAG.  Consider two vertices $u,v \in V$ such that there is a directed path from $u$ to $v$ in $G$ and consequently no path from $v$ to $u$ because $G$ is a DAG.  We define two policies $\pi=\curly{u_1,...,u,v,...,u_n}$ and $\pi'=\curly{u_1,...,v,u,..., u_n}$.   The only difference in the two policies is that $u$ and $v$ are swapped.  In $\pi$ vertices $u$ and $v$ respect the partial order of $G$, whereas in $\pi'$ they do not.    We will now show that the expected follows of $\pi'$ is less than  $\pi$.   

The expected follows of a policy $\pi$ can be written as
$\sum_{w\in V}\mathbf E\bracket{ X_w|\pi}$, where $X_w$ is one if the agent interaction with $w$ results in a follow by $w$ and zero otherwise.  We want to show that the expected follows will be larger if $u$ is followed before $v$.  For any vertex $w$ we define $\delta_w=\mathbf E\bracket{X_{w}|\pi}-\mathbf E\bracket {X_{w}|\pi'}$ and we define the difference in the expected follows of the two policies as $\delta = \sum_{w\in V}\delta_w$.  For any vertex $w$ which occurs before  $v$ and $u$  in either policy, we have that $\delta_w=0$.  This is because the state of the graph when the agent interacts with these vertices is the same under each policy.  For vertex $u$ we have $\delta_u=0$ because $u$ does not follow $v$ and there is no path from $v$ to $u$.  Therefore, there is no way for $v$ to affect the overlap of $u$ with the agent, and thus no way to affect the subsequent follow probability.  For $v$ and all vertices after $u$ and $v$ in the policies, we use the following lemma.

\begin{lemma}\label{lem:path}
	Let $(u_1,u_2,...,u_n)$ be a path on a directed graph $G$.  Then $\mathbf E[X_{u_j}|X_{u_i}=1] > \mathbf E[X_{u_j}|X_{u_i}=0]$ for $1\leq i<j\leq n$.
\end{lemma}
This lemma shows that if the agent can get a vertex $w$ to follow it, it will receive a boost in the expected follows for any vertex that is a descendant of $w$. 

 Now consider vertex $v$, which has no path to $u$, but is a descendant of $u$.  Then we have   
\begin{align*}
\delta_{v} =& \mathbf E\bracket{X_{v}|X_{u}=1,\pi}\mathbf E\bracket{X_{u}|\pi}\\
&+\mathbf E\bracket{X_{v}|X_{u}=0,\pi}(1-\mathbf E\bracket{X_{u}|\pi})\\
&-\mathbf E\bracket{X_{v}|X_{u}=0,\pi'}\\
=&\mathbf E\bracket{X_{u}|\pi}(\mathbf E\bracket{X_{v}|X_{u}=1,\pi}-\mathbf E\bracket{X_{v}|X_{u}=0,\pi})\\	
>&0.
\end{align*}
Above we have used Lemma \ref{lem:path} along with the fact that $X_i$ are Bernoulli random variables, so $\mathbf E[X_{i}]=\mathbf P(X_{i}=1)$.  We also made use of the fact that $\mathbf E\bracket{X_{v}|X_{u}=0,\pi}=\mathbf E\bracket{X_{v}|X_{u}=0,\pi'}$ because if $u$ does not follow the agent, the conditional follow probability is the same under both policies when the agent interacts with $v$.

For any vertex $w$ that occurs after $u$ and $v$ in the policies, we have two situations.  Either there is no path from $v$ to $w$, in which case $\delta_w=0$ because $v$ cannot affect the follow probability of $w$, or there is such a path.  In the case where a path exists, we have
\begin{align*}
\delta_{w} =& \mathbf E\bracket{X_{w}|X_{v}=1,\pi}\mathbf E\bracket{X_{v}|\pi}
+\mathbf E\bracket{X_{w}|X_{v}=0,\pi}(1-\mathbf E\bracket{X_{v}|\pi})\\
&-\mathbf E\bracket{X_{w}|X_{v}=1,\pi'}\mathbf E\bracket{X_{v}|\pi'}
-\mathbf E\bracket{X_{w}|X_{v}=0,\pi'}(1-\mathbf E\bracket{X_{v}|\pi'})\\
=&\mathbf E\bracket{X_{w}|X_v=1}(\mathbf E\bracket{X_{v}|\pi}-\mathbf E\bracket{X_{v}|\pi'})
+\mathbf E\bracket{X_{w}|X_v=0}(\mathbf E\bracket{X_{v}|\pi'}-\mathbf E\bracket{X_{v}|\pi})\\
=& (\mathbf E\bracket{X_{v}|\pi}-\mathbf E\bracket{X_{v}|\pi'})
(\mathbf E\bracket{X_{w}|X_v=1}-\mathbf E\bracket{X_{w}|X_v=0})\\
=&\delta_v(\mathbf E\bracket{X_{w}|X_v=1}-\mathbf E\bracket{X_{w}|X_v=0})\\
>&0.
\end{align*}
Here, in addition to Lemma \ref{lem:path} and our result for $\delta_v$, we also used the fact that the expected value of $X_w$ conditioned on $X_v$ is the same for both policies because the relevant graph state is the same.
This result shows that $\delta >0$, and $\pi$ has more expected follows than $\pi'$.  Therefore, any policy can increase its expected follows by swapping any pair of adjacent vertices so they respect the partial order of the underlying DAG.  This process can be continued until the policy is a linear extension of the DAG and no further increase in the expected follows is possible.

\section{Proof of Theorem \ref{thm:dag1}}\label{app:thmdag1}
We require the following result which is  for 
follow probability functions that are non-decreasing in the overlap.
\begin{lemma}\label{lem:parents}
	Consider a a DAG $G=(V,E)$ and a vertex $v\in V$ with parents given by the set $P(v)\subset V$, $|P(v)|=n$.  Let the follow probability for $v$ with feature set $F_v$ and overlap $\phi$ be given by $p(\phi,F_v) = g(F_v)f(\phi)$, where $f$ is a non-decreasing function.  Let $p_v$ be the follow probability of $v$ under a linear extension policy.   Then
	we have that
	\begin{align}
	p_v = g(F_v)\sum_{k=0}^{n}\Delta_k\sum_{S\subseteq P(v): |S|=k}\prod_{u\in S}p_u \label{eq:parents}
	\end{align}
	where we define $\Delta_k = \sum_{i=0}^ka^k_i f(i)$, where $a^k_i = a^{k-1}_{i-1}-a^{k-1}_i$, $a^k_{k+1}=0$, $a^k_k=1$, $a^k_0 = (-1)^k$, and $a^k_{-1}=0$.
\end{lemma}
If we assume that $f(\phi)$ is linear, then equation \eqref{eq:parents} takes a simpler form.  
We use the following result regarding the terms $a^k_i$.
\begin{lemma}\label{cor:sum}
	Let the terms $a^k_i$ be given in Lemma \ref{lem:parents}.  Then we have that 
	$	\sum_{i=0}^k a^k_i = 0$ for $k\geq 0$, $\sum_{i=0}^1a^1_ii = 1$, and 
	$\sum_{i=0}^ka^k_ii = 0$ for $k > 1$.
\end{lemma}
For $f(\phi)=1+\beta\phi$, the terms $\Delta_k$ take the following form.
\begin{align*}
\Delta_k &= \sum_{i=0}^ka^k_i+\beta\sum_{i=0}^ka^k_ii
\end{align*}
Using Lemma \ref{cor:sum} we find that
\begin{align*}
\Delta_0 &= 1\\
\Delta_1 &= \beta\\
\Delta_k &= 0,~~k\geq 2.	
\end{align*}
Substituting this into equation \eqref{eq:parents} we obtain
\begin{align}
p_v & = g(F_v)\paranth{1+\beta\sum_{u\in P(v)}p_u}. \label{eq:parents_linear}
\end{align}
This expression is a recursion in the depth of the DAG.  We use this to prove the result
by induction on the distance of a vertex from the root of a DAG.
The base case is for a vertex $u$ which is a root and has no parents.  We obtain
$p_u = g(F_u)$, which agrees with equation \eqref{eq:dag1}. 

 Now consider a vertex $v$ which is not a root in a DAG and assume equation \eqref{eq:dag1} is true for all of its parents, which we denote $P(v)$. Also assume there are $N$ vertices in the DAG.  From equation \eqref{eq:parents_linear} we have 
\begin{align*}
p_v &= g(F_v)+\beta g(F_v)\sum_{u\in P(v)}p_u\\
    &= g(F_v)+\beta g(F_v)\sum_{u\in P(v)}\sum_{l=0}^{N-1}\beta^l\sum_{P\in\mathcal P_l(u,G)}\prod_{w\in P}g(F_w) \\
    & = \sum_{l=0}^{N-1}\beta^l\sum_{P\in\mathcal P_l(v,G)}\prod_{u\in P}g(F_u).
\end{align*}
In the last line above, the sum changes from paths that terminate on
parents of $v$ to paths that terminate on $v$. As a result,
we recover equation \eqref{eq:dag1}.
\section{Proof of Lemma \ref{lem:path}}

We will prove the result by induction.    For any $1<i<n$ we have that
$\mathbf E[X_{u_{i+1}}|X_{u_i}=1]>\mathbf E[X_{u_{i+1}}|X_{u_i}=0]$ because the follow probabilities are monotonically increasing in the overlap and $u_{i+1}$ follows $u_i$.  This gives our base case of $\mathbf E[X_{u_{2}}|X_{u_1}=1]>\mathbf E[X_{u_{2}}|X_{u_1}=0]$.  We then assume that $\mathbf E[X_{u_i}|X_{u_1}=1]>\mathbf E[X_{u_i}|X_{u_1}=0]$ for $2<i<n$.  For the case of $i+1$ we then have that for $k \in\curly{0,1}$
\begin{align*}
\mathbf E[X_{u_{i+1}}|X_{u_1}=k] =&  \mathbf E[X_{u_{i+1}}|X_{u_i}=1]\mathbf E[X_{u_{i}}|X_{u_1}=k]+\\
&\mathbf E[X_{u_{i+1}}|X_{u_i}=0](1-\mathbf E[X_{u_{i}}|X_{u_1}=k])\\
=& \mathbf E[X_{u_{i+1}}|X_{u_i}=0]+\\
& \mathbf E[X_{u_{i}}|X_{u_1}=k](\mathbf E[X_{u_{i+1}}|X_{u_i}=1]-\\
&\mathbf E[X_{u_{i+1}}|X_{u_i}=0]).
\end{align*}
Now we define $\delta = \mathbf E[X_{u_{i+1}}|X_{u_1}=1] -\mathbf E[X_{u_{i+1}}|X_{u_1}=0]$.  Using the above result we have
\begin{align*}
\delta =& (\mathbf E[X_{u_{i+1}}|X_{u_i}=1]-\mathbf E[X_{u_{i+1}}|X_{u_i}=0])(\mathbf E[X_{u_{i}}|X_{u_1}=1]-\mathbf E[X_{u_{i}}|X_{u_1}=0])\\
&>0.
\end{align*}
Above we used the induction hypothesis and the monotonicity of the follow probabilities in the overlap.
\subsection{Proof of Lemma \ref{lem:parents}}\label{sec:parents}

We prove the result by induction on the number of parents of a vertex.  The induction hypothesis is given by equation \eqref{eq:parents}.  We assume that $v$ has $n$ parents.  For $n=0$, we have $p_v = f(0)g(F_v)$, which is just the follow probability of a vertex with no overlap with the agent. 

Next, we assume equation \eqref{eq:parents} is true for a vertex $v$ with $n$ parents.
Now consider adding the $n+1$ parent node.  We denote this parent vertex as $u_{n+1}$   If this vertex does not follow the agent,
$p_v$ is not changed. If it
follows the agent, the overlap increases by one and the possible values for the overlap of $v$ with the agent is lower bounded by one.  This means that the value of $p_v$, conditioned on this event, is
given by equation \eqref{eq:parents}, except that the argument of $f$ is increased by one.  This shifts the values of the $\Delta_k$ terms.
To simplify our notation, we define $\Delta_k' = \sum_{i=0}^{k}a^k_i f(i+1)$.  We let $P(v)$ be the parents of $v$ not including $u_{n+1}$.  Then we have
\begin{align}
p_v =& \mathbf E[X_v|X_{u_{n+1}}=0](1-p_{u_{n+1}})+\mathbf E[X_v|X_{u_{n+1}}=1]p_{u_{n+1}}\nonumber\\
=&(1-p_{u_{n+1}})g(F_v)\sum_{k=0}^{n}\Delta_k\sum_{S\subseteq P(v): |S|=k}\prod_{u\in S}p_u\nonumber\\
&+p_{u_{n+1}}g(F_v)\sum_{k=0}^{n}\Delta'_k\sum_{S\subseteq P(v): |S|=k}\prod_{u\in S}p_u\nonumber\\
=&g(F_v)\sum_{k=0}^{n}\Delta_k\sum_{S\subseteq P(v): |S|=k}\prod_{u\in S}p_u\nonumber\\
&+p_{u_{n+1}}g(F_v)\sum_{k=0}^{n}(\Delta'_k-\Delta_k)\sum_{S\subseteq P(v): |S|=k}\prod_{u\in S}p_u\label{eq:delta}
\end{align}
The difference $\Delta'_k-\Delta_k$ is given by
\begin{align*}
\Delta'_k-\Delta_k & = \sum_{i=0}^ka_i^k(f(i+1)-f(i))\\
& = a^k_kf(k+1)-a^k_kf(k) +a^k_{k-1}f(k)-a^k_{k-1}f(k-1)+... +a^k_0f(1)-a^k_0f(0)\\
& = f(k+1)(a^k_k -a^k_{k+1})+f(k)(a^k_{k-1} -a^k_{k})+...+f(0)(a^k_0 -a^k_{-1})\\
& = \sum_{i=0}^{k+1}a^{k+1}_if(i)\\
& = \Delta_{k+1}.
\end{align*}
Above we used the definition of $a^k_i$ given in Lemma \ref{lem:parents}.    Substituting this into equation \eqref{eq:delta} and letting $P'(v)$ denote the parents of $v$ including $u_{n+1}$, we obtain
\begin{align}
p_v = & g(F_v)\sum_{k=0}^{n}\Delta_k\sum_{S\subseteq P(v): |S|=k}\prod_{u\in S}p_u + p_{u_{n+1}} g(F_v)\sum_{k=0}^{n}\Delta_{k+1}\sum_{S\subseteq P(v): |S|=k}\prod_{u\in S}p_u\nonumber\\
=& g(F_v)\paranth{\sum_{k=0}^{n}\Delta_k\sum_{S\subseteq P(v): |S|=k}\prod_{u\in S}p_u +\sum_{k=0}^{n}\Delta_{k+1}\sum_{S\subseteq P(v): |S|=k}p_{u_{n+1}}\prod_{u\in S}p_u}\nonumber\\
=& g(F_v)\sum_{k=0}^{n+1}\Delta_k\sum_{S\subseteq P'(v): |S|=k}\prod_{u\in S}p_u\nonumber.
\end{align}
The above expression matches the induction hypothesis, completing the proof.

\section{Proof of Lemma \ref{cor:sum}}
The results are proved using the properties of the $a^k_i$ terms given
in Lemma \ref{lem:parents}
We first prove the result for $\sum_{i=0}^{1}a^1_ii$.
\begin{align*}
\sum_{i=0}^{1}a^1_ii & = 0+a^1_1 = 1.
\end{align*}
Next we prove the result for $\sum_{i=0}^{k}a^k_i$
\begin{align*}
\sum_{i=0}^{k}a^k_i & = \sum_{i=0}^{k}(a^{k-1}_{i-1}-a^{k-1}_i)\\
& = a^{k-1}_{-1}-a^{k-1}_k = 0.
\end{align*}
Finally, we prove the result for $\sum_{i=0}^{k}a^k_ii$
\begin{align*}
\sum_{i=0}^{k}a^k_ii & = \sum_{i=0}^{k}(a^{k-1}_{i-1}-a^{k-1}_i)i\\
& =0+\sum_{i=0}^{k-1} a^{k-1}_{i}  -a^{k-1}_kk\\
& = 0.
\end{align*}
Above we used the result $\sum_{i=0}^{k}a^k_i i=0$.
%

\bibliographystyle{plainnat}
\bibliography{main}

\end{document}